\documentclass[a4paper,10.5pt]{article}
\usepackage{theorem}
\usepackage{latexsym,amssymb,amsfonts,amsmath,cite}
\usepackage[dvips]{graphicx}
\usepackage{epsfig}
\usepackage{caption}
\newtheorem{theorem}{Theorem}
\newtheorem{lemma}{Lemma}

\title{{\Large {\bf Weak convergence of the Wojcik model 
}
}}

\author{ 
{\small 
Takako Endo,$^{1}$ 
\footnote{endo.takako@ocha.ac.jp 
}\quad Norio Konno,$^{2}$ 
\footnote{konno@ynu.ac.jp
}\quad  
}\\ 
{\scriptsize $^{1}$ 
Department of Physics, Ochanomizu University,
}\\
{\scriptsize Bunkyo, Tokyo, 112-0012, Japan
} \\
{\scriptsize $^2$ 
Department of Applied Mathematics, Faculty of Engineering, Yokohama National University,
}\\
{\scriptsize Hodogaya, Yokohama, 240-8501, Japan
} \\
} 

\date{\empty }
\begin{document}
\maketitle

\par\noindent
\begin{small}
\baselineskip=24pt
\par\noindent
{\bf Abstract}

We study ``the Wojcik model'' which is a discrete-time quantum walk (QW) with one defect in one dimension, introduced by Wojcik et al..  For the Wojcik model, we give the weak convergence theorem describing the ballistic behavior of the walker in the probability distribution in a rescaled position-space.
In our previous studies, we obtained the time-averaged limit and stationary measures concerning localization for the Wojcik model. As a result, we get the mathematical expression of the whole picture of the behavior of the walker for the Wojcik model.
Here the coexistence of localization and the ballistic spreading is one of the peculiar properties of one-dimensional QWs with one defect. Due to the coexistence, it has been strongly expected to utilize QWs to quantum search algorithms.
In order to derive the weak convergence theorem, we take advantage of the generating function method. 
We emphasize that the time-averaged limit measure is symmetric for the origin, however, the weight function in the weak limit measure is asymmetric in general, which implies that the weak convergence theorem represents the asymmetry of the probability distribution. Furthermore, the weak limit measure heavily depends on the phase of the defect and initial state of the walker.
Comparing with our previous studies, we also show some numerical results of the probability distribution to confirm that our result is relevant mathematically, and consider the effect of changing the phase and initial coin state on the probability distribution, or the ballistic spreading, which is one of the motivations of our study.

\end{small}

\baselineskip=24pt

\setcounter{equation}{0}
\section{Introduction}

This paper is the sequential to \cite{watanabe, endo}. 
Quantum walks (QWs) are quantum counter parts of classical random walks, and have been intensively studied in various fields, such as computer science \cite{ambainis,eli} and quantum physics \cite{kitagawa, mol}.  
Owing to the rich applications, it is worth to study QWs both theoretically and experimentally.
Especially, it is very important to study the behavior of the walker in the long-time limit.  Here, the time evolution of QWs are defined by unitary evolutions of probability amplitudes. On the other hand, classical random walks are obtained by
evolutions of probabilities by transition matrices. 
Recently, QWs have been also implemented experimentally by various materials, such as trapped ions \cite{zari} and photons \cite{peret}. 
However, we have not been able to experimentally grasp the behavior in the long-time limit, since it is too difficult for the QWs to implement the state after many steps. Moreover, because of its quantumness, it is hard to intuitively understand the properties of QWs. \\
\indent
As recent studies of QWs suggested, the QWs in one dimension, show two characteristic behaviors in the long-time limit, that is, ``localization''and ``the ballistic spreading''. 
In detail, some of the
quantum walkers may localize and return to the starting point even in the long-time limit, which is in
marked contrast to the classical random walks which do not show such localized behaviors. 
Furthermore, the quantum walker spreads much faster than the classical one. Indeed, for the QWs, the width
of the probability distribution diverges with the order of time $t$. On the other hand, the classical random
walker diverges with the order of $\sqrt{t}$. Due to the coexistence of localized and the ballistic behavior, the QWs are believed to be far more efficient for search algorithms
than the classical random walks, since the quantum walker spreads much faster than the classical one and can localize at the target.\\
\indent
Up to this day, two kinds of limit theorems describing the behavior of discrete-time QWs in one dimension have been constructed \cite{segawa}.
One is the limit theorem concerning localization, and the other is the limit theorem concerning the ballistic behavior of the quantum walker.
For the mathematical aspects of the QWs, Konno et al. \cite{segawa} introduced three kinds of measures for one-dimensional QWs: time-averaged limit measure, weak limit measure, and stationary measure. The first two
measures describe a coexistence of localized and the ballistic behavior in the QW, respectively.  In this paper, we focus on the weak limit measure. Now we assume that $X_{t}$ is a discrete-time QW at time $t$. Then, the weak limit measure of $X_{t}/t$ is described in general as follows; There exist a rational polynomial $w(x)$, and $C\in[0,1),\;a\in(0,1)$ such that 
\begin{align}\mu(dx)=C\delta_{0}(dx)+w(x)f_{K}(x;a)dx\label{weakmeasure}\end{align}
where
\begin{align}f_{K}(x;a)=\dfrac{\sqrt{1-a^{2}}}{\pi(1-x^{2})\sqrt{a^{2}-x^{2}}}I_{(-a,a)}(x)\label{konnofunction}\end{align}
with
\begin{eqnarray*}I_{A}(x)=\left\{ \begin{array}{ll}
1&(x\in A)\\
0&(x\notin A)
\end{array} \right..
\end{eqnarray*}  
We should note that $f_{K}(x;a)dx$ is the weak measure for the Hadamard walk \cite{konnoweak}. 
By recent studies of QWs, it is strongly expected that most of the discrete-time QWs have the weak limit measure written by the convex combination of the term of delta function $\delta_{0}(dx)$ and the absolutely continuous part $w(x)f_{K}(x;n)dx$. We therefore, expect the existence of the universality class for the QWs having the weak limit measure expressed by Eq. \eqref{weaklimit} \cite{segawa}.  

In this paper, we study ``the Wojcik model'', introduced by Wojcik et al. \cite{wojcik}. By the numerical results in main, they reported that changing a phase at a single point gives astonishing localization effect. Then Endo et al. derived the mathematical forms of localization, that is, the stationary and time-averaged limit measures, and they showed the astonishing localization effect mathematically \cite{watanabe, endo}. Based on the previous results, we focus on the ballistic behavior of the Wojcik model to clarify the whole picture of the asymptotic behavior. As a result, we obtain the weak convergence theorem, the mathematical expression of the ballistic spreading. We also give numerical results of the probability distributions for some phase parameters of the defect and initial coin states, and then consider what the weak convergence theorem suggests.

\indent
The rest of this paper is organized as follows. In Section \ref{model},
we define the Wojcik model which is the main target in this paper, and present our main result, the weak convergence theorem of the Wojcik model. 
Then in Section \ref{examples}, we show the numerical results of the probability distribution for some phase parameters and initial coin states, and consider what our analytical result implies.
Appendix A is devoted to the proof of Theorem \ref{weaklimit}.
\section{Model and main result}
\label{model}
\subsection{Definition of a discrete-time QW: the Wojcik model}
Let us introduce the total space of discrete-time QW, ${\cal H}$, which is a Hilbert space consisting of two Hilbert spaces ${\cal H}_{C}$ and ${\cal H}_{P}$, that is,
\[{\cal H}={\cal H}_{P}\otimes{\cal H}_{C},\] 
where 
\[{\cal H}_{P}=Span\{|x\rangle; x\in\mathbb{Z}\},\quad{\cal H}_{C}=Span\{|J\rangle; J\in\{L,R\}\},\]
with
\[|L\rangle=\begin{bmatrix}1 \\ 0\end{bmatrix},\quad |R\rangle=\begin{bmatrix}0 \\ 1\end{bmatrix}.\]
We should note that ${\cal H}_{C}$ and ${\cal H}_{P}$ represent the position and the direction of the motion of the walker, respectively.
In general, discrete-time QW has a state at each time $t$ and position $x$, called ``qubit'' written by a two-dimensional vector
\[\Psi_{t}(x)=\begin{bmatrix}\Psi^{L}_{t}(x)\\ \Psi^{R}_{t}(x) \end{bmatrix}\in\mathbb{C}^{2},\]
and we can define the state of the system at each time $t$ by
\[\Psi_{t}=[\cdots,\Psi_{t}(-1),\Psi_{t}(0),\Psi_{t}(1),\cdots]\in(\mathbb{C}^{2})^{\mathbb{Z}}.\]
In this paper, we focus on a discrete-time QW with
one defect on the line, whose time evolution is defined by the unitary matrices on ${\cal H}_{C}$ as follows; 
\begin{align}
U_{x}=\left\{ \begin{array}{ll}
\dfrac{1}{\sqrt{2}}\begin{bmatrix}
1 & 1\\
1 & -1\\
\end{bmatrix}&(x\neq0 ),\\
\\
\dfrac{e^{2\pi i\phi}}{\sqrt{2}}\begin{bmatrix}
1 & 1\\
1 & -1\\
\end{bmatrix}&(x=0),
\end{array} \right.\label{wojcik_def}
\end{align}
\par\noindent
where $\phi\in[0,1)$. Here $U_{x}$ is called ``the quantum coin''.
To consider the time evolution, we divide the unitary matrices into $P_{x}$ and $Q_{x}$ as 
\begin{align*}
P_x = \left\{ \begin{array}{ll}
\dfrac{1}{\sqrt{2}}\begin{bmatrix}
1 & 1\\
0 & 0\\
\end{bmatrix}&(x\neq0 ),\\
\\
\dfrac{e^{2\pi i\phi}}{\sqrt{2}}\begin{bmatrix}
1 & 1\\
0 & 0\\
\end{bmatrix}&(x=0),
\end{array} \right.
\qquad 
Q_x = \left\{ \begin{array}{ll}
\dfrac{1}{\sqrt{2}}\begin{bmatrix}
0 & 0\\
1 & -1\\
\end{bmatrix}&(x\neq0 ),\\
\\
\dfrac{e^{2\pi i\phi }}{\sqrt{2}}\begin{bmatrix}
0 & 0\\
1 & -1\\
\end{bmatrix}&(x=0),
\end{array} \right.
\end{align*}
\par\noindent
with $U_x = P_x + Q_x$. 
Here, $P_x$ and $Q_x$ are equivalent to the left and right movements, respectively. Using operators $P_{x}$ and $Q_{x}$,
the time evolution is determined by the recurrence formula;
\begin{align*}
\Psi_{t+1} (x) = P_{x+1} \Psi_t (x+1) + Q_{x-1} \Psi_t (x-1) \quad (x \in \mathbb{Z}).
\end{align*}
In this paper, we call the QW ``the Wojcik model''.
We studied localization for our Wojcik model in \cite{watanabe, endo}, that is, we derived the time-averaged limit and stationary measures for the Wojcik model which describe localization mathematically.
Therefore, we obtain the mathematical description of the whole picture of the motion of the Wojcik model in the long-time limit by the weak convergence theorem describing the ballistic behavior.

\subsection{Main result: Weak convergence theorem}
Let $X_{t}$ be the quantum walker at time $t$, and we introduce the characteristic function of $X_{t}/t$,
\[E\left[e^{i\xi\frac{X_{t}}{t}}\right]=\sum_{j\in\mathbb{Z}}P(X_{t}=j)e^{i\xi\frac{X_{t}}{t}},\]
where $P(X_{t}=j)$ is the probability that $X_{t}=j$ holds.
In this subsection, we consider the expression of $E[e^{i\xi X_{t}/t}]$ in the long-time limit $t\to\infty$.
According to \cite{segawa}, we see
\[1=\left.\left(\lim_{t\to\infty}E\left[e^{i\xi\frac{X_{t}}{t}}\right]\right)\right|_{\xi=0}=C+\int_{-\infty}^{\infty}w(x)f_{K}(x;1/\sqrt{2})dx,\]
with \[C=\sum_{x}\overline{\mu}_{\infty}(x).\]
Here, we should note that $\overline{\mu}_{\infty}(x)$ is the time-averaged limit measure describing localization.\\
\indent
From now on, we give the weak convergence theorem for the missing part $1-C$ with $0\leq C< 1,$ which describes in general the ballistic behavior of QW \cite{konnoweak}. 
\par\indent
\begin{theorem}
\label{weaklimit}
Assume that the Wojcik model starts from the origin with the initial coin state $\Psi_{0}(0)={}^T\![\alpha,\beta]$, where $\alpha,\beta\in\mathbb{C}$. 
Put $\alpha=ae^{\phi_{1}},\;\beta=be^{\phi_{2}}$ with $a,b\geq0,\;a^{2}+b^{2}=1$ and $\phi_{1}, \phi_{2}\in\mathbb{R}$, where $\mathbb{R}$ is the set of real numbers. 
Let $\tilde{\phi}_{12}=\phi_{1}-\phi_{2}\;$.
For the Wojcik model, $X_{t}/t$ converges weakly to the random variable $Z$ which has the following probability density function;
\begin{eqnarray}\mu(x)=C\delta_{0}(x)+w(x)f_{K}(x;1/\sqrt{2})
,\label{siki}\end{eqnarray}
where 
\begin{align*}f_{K}(x;1/\sqrt{2})=\dfrac{1}{\pi(1-x^{2})\sqrt{1-2x^{2}}}I_{(-1/\sqrt{2},1/\sqrt{2})}(x)\end{align*}
with
\begin{eqnarray*}I_{D}(x)=\left\{ \begin{array}{ll}
1&(x\in D)\\
0&(x\notin D)
\end{array} \right..
\end{eqnarray*} 
Here, the weight function $w(x)$ is given by
\begin{eqnarray}
w(x)=\dfrac{t_{3}x^{5}+t_{2}x^{4}+t_{1}x^{3}+t_{0}x^{2}}{s_{2}x^{4}+s_{1}x^{2}+s_{0}},
\end{eqnarray}
where
\begin{eqnarray*}s_{2}\!\!\!\!\!&=&\!\!\!\!\!\cos^{2}(4\pi\phi),\;s_{1}\!\!=\!\!8\sin^{2}(\pi\phi)(\cos(4\pi\phi)+4\sin^{2}(\pi\phi)\sin^{2}(2\pi\phi)),\;s_{0}\!\!=\!\!16\sin^{4}(\pi\phi)\cos^{2}(2\pi\phi),\\
t_{3}\!\!\!\!\!\!&=&\!\!\!\!\!\!\left\{ \begin{array}{ll}
a_{2}\cos(4\pi\phi) & (x\geq0) \\
-b_{2}\cos(4\pi\phi) & (x<0) \\
\end{array} 
\right.,\quad t_{2}\!=\!\left\{ \begin{array}{ll}
a_{1}\cos(4\pi\phi)+8a_{3}\sin^{2}(\pi\phi)\sin(2\pi\phi) & (x\geq0) \\
b_{1}\cos(4\pi\phi)+8b_{3}\sin^{2}(\pi\phi)\sin(2\pi\phi) & (x<0) \\
\end{array} 
\right.,\\
t_{1}\!\!\!\!\!\!&=&\!\!\!\!\!\!\left\{ \begin{array}{ll}
4a_{2}\sin^{2}(\pi\phi) & (x\geq0)\\
-4b_{2}\sin^{2}(\pi\phi) & (x<0)\\
\end{array} 
\right.,\quad
t_{0}\!=\!\left\{ \begin{array}{ll}
-4\sin^{2}(\pi\phi)(a_{3}\sin(2\pi\phi)-a_{1}) & (x\geq0)\\
-4\sin^{2}(\pi\phi)(b_{3}\sin(2\pi\phi)-b_{1}) & (x<0)\\
\end{array} 
\right.,
\end{eqnarray*}
with
\begin{eqnarray*}
\left\{
\begin{array}{l}
a_{1}=1+2a^{2}-2ab\cos\tilde{\phi}_{12}-2a^{2}\cos(2\pi\phi)+2ab\cos(\tilde{\phi}_{12}+2\pi\phi),\\ a_{2}=1-2a^{2}-2ab\cos\tilde{\phi}_{12},\\ 
a_{3}=2a(a\sin(2\pi\phi)-b\sin(\tilde{\phi}_{12}+2\pi\phi)),
\end{array}
\right.\end{eqnarray*}
and
\begin{eqnarray*}
\left\{
\begin{array}{l}
b_{1}=1+2b^{2}+2ab\cos\tilde{\phi}_{12}-2ab\cos(\tilde{\phi}_{12}-2\pi\phi)-2b^{2}\cos(2\pi\phi),\\
b_{2}=1-2b^{2}+2ab\cos\tilde{\phi}_{12},\\
 b_{3}=2b(-a\sin(\tilde{\phi}_{12}-2\pi\phi)+b\sin(2\pi\phi)).
\end{array}
\right.\\
\end{eqnarray*}
Note that $\phi$ is defined by Eq. \eqref{wojcik_def}.
\end{theorem}
We should remark that $f_{K}(x;1/\sqrt{2})$ is the density function of the Hadamard walk in a rescaled position-space \cite{konnoweak}. 
Moreover, the second term of Eq. \eqref{siki}, $w(x)f_{K}(x;1/\sqrt{2})$, is an absolutely continuous part.
The proof of Theorem \ref{weaklimit} is given in Appendix A.

We emphasize that the weak limit measure heavily depends on the phase parameter of the defect and the initial coin state. 
We also note that the weight function is generally asymmetric for the origin, however, it can be symmetric with appropriate choice of the phase parameter and initial coin state. The asymmetry of the weight function correponds to that of the probability distribution. As concrete examples, we will show the numerical results of the probablity distributions for some phases and initial coin states in the next section. As we see in Section \ref{examples}, it can be expected that the symmetric initial coin states and appropriate choice of the phase parameter will contribute to the symmetry of probability distributions.

\section{Examples}
\label{examples}
We consider the Wojcik model for some phase parameters of the defect and initial coin states as follows;

\begin{enumerate}
\item\underline{{\bf Case of the Hadamard walk:}}\\
First of all, we see the Hadamard walk whose quantum coin is given by
\begin{align}
U_{x}=
\dfrac{1}{\sqrt{2}}\begin{bmatrix}1 & 1\\ 1 & -1 \end{bmatrix}. \label{example1}\end{align}
\par\noindent
The Hadamard walk can be obtained by putting $\phi=0$ in Eq. \eqref{wojcik_def}.
\begin{enumerate}
\item 
Put the initial coin state $\Psi_{0}(0)={}^T\![1,0]$.
Theorem \ref{weaklimit} gives the weight function $w(x)$ in Eq. \eqref{weakmeasure} by
\[w(x)=1-x.\]
Hence, we have
\begin{align}\int_{-\frac{1}{\sqrt{2}}}^{\frac{1}{\sqrt{2}}}w(x)f_{K}(x;1/\sqrt{2}) dx=1.\end{align}

\item 
Let the initial coin state be $\Psi_{0}(0)={}^T\![i/\sqrt{2},1/\sqrt{2}]$.
In a similar way, we obtain the weight function $w(x)$ by
\[w(x)=1.\]
Therefore, we get
\begin{align}\int_{-\frac{1}{\sqrt{2}}}^{\frac{1}{\sqrt{2}}}w(x)f_{K}(x;1/\sqrt{2}) dx=1.\end{align}
\par\indent
\par\noindent
From the above, we obtain the same weight functions as the previous studies \cite{konnoqw, konnoweak} from our result. 
Here we should note that the Hadamard walk does not localize in long time limit \cite{konnoqw, konnoweak}, and we see that $C=0$ in Eq. \eqref{weakmeasure}.

\end{enumerate}
\item \underline{{\bf QW with one defect: $\phi=1/2$ case.}}\\
We consider the QW whose quantum coin is given by
\begin{align}
U_{x}=\left\{ \begin{array}{ll}
\dfrac{1}{\sqrt{2}}\begin{bmatrix}1 & 1\\ 1 & -1 \end{bmatrix}\quad (x=\pm1,\pm2,\cdots), \\
\\
\dfrac{1}{\sqrt{2}}\begin{bmatrix}-1 & -1\\ -1 & 1 \end{bmatrix} \quad (x=0),\\
\end{array}\right. \label{example3}\end{align}
\par\noindent
which is obtained by putting $\phi=1/2$ in Eq. \eqref{wojcik_def}.
\begin{enumerate}
\item 
Let the initial coin state be $\Psi_{0}(0)={}^T\![1,0]$.
Theorem \ref{weaklimit} gives the weight function $w(x)$ in Eq. \eqref{siki} by
\[w(x)=\left\{ \begin{array}{ll}
\dfrac{-x^{3}+5x^{2}}{x^{2}+4} & (x\geq0), \\
\dfrac{-x^{3}+x^{2}}{x^{2}+4} & (x<0).
\end{array} \right.\]
Hence, we see
\begin{align}\int_{-\frac{1}{\sqrt{2}}}^{\frac{1}{\sqrt{2}}}w(x)f_{K}(x;1/\sqrt{2}) dx=\dfrac{1}{5}.\end{align}
Now, we should note that we obtained the time-averaged limit measure $\overline{\mu}_{\infty}(x)$ by Theorem $2$ in \cite{endo}, and as a result, we obtain the coefficient of the delta function $\delta_{0}(x)$ in Eq. \eqref{siki} by
\begin{align}C=\sum_{x}\overline{\mu}_{\infty}(x)=\dfrac{8}{25}+2\times\dfrac{12}{25}\sum_{y=1}^{\infty}\left(\dfrac{1}{5}\right)^{y}=\dfrac{4}{5},
\end{align}
since
\[
\left\{ \begin{array}{l}
\overline{\mu}_{\infty}(0)=\dfrac{8}{25},\\
\overline{\mu}_{\infty}(x)=\dfrac{12}{25}\left(\dfrac{1}{5}\right)^{|x|}.
\end{array} \right.\]
\noindent
Therefore, we have
\[C+\int_{-\frac{1}{\sqrt{2}}}^{\frac{1}{\sqrt{2}}}w(x)f_{K}(x;1/\sqrt{2}) dx=1.\]\\
Here, we give the numerical results of the probability distribution at time $t=100, 1000$, and $10000$ in re-scaled spaces $(x/t, tP_{t}(x))\;(t=100,1000,10000)$, where $x$ represents the position of the walker and $P_{t}(x)$ is the probability that the walker exists on position $x$ at time $t$. We should remark that $x/t$ corresponds to the real axis, and $tP_{t}(x)$ corresponds to the imaginary axis, respectively. Also, we put the graph of $w(x)f_{K}(x;1/\sqrt{2})$, which is related to absolutely continuous part of the weak limit measure $\mu(dx)$, on the picture at each time. We see that the graph of $w(x)f_{K}(x;1/\sqrt{2})$ is right on the middle of the probability distribution for each position at each time, which suggests that our result is mathematically proper. 
We also emphasize that $\overline{\mu}_{\infty}(x)$ is symmetric for the origin \cite{endo}, however, $w(x)f_{K}(x;1/\sqrt{2})$ does not have an origin symmetry (Figs. 1,3,5) , which suggests that the weak limit measure represents the asymmetry of the probability distribution (Figs. 1,3,5).

\begin{figure}[h]
\begin{minipage}{0.5\hsize}
\centerline{\epsfig{file=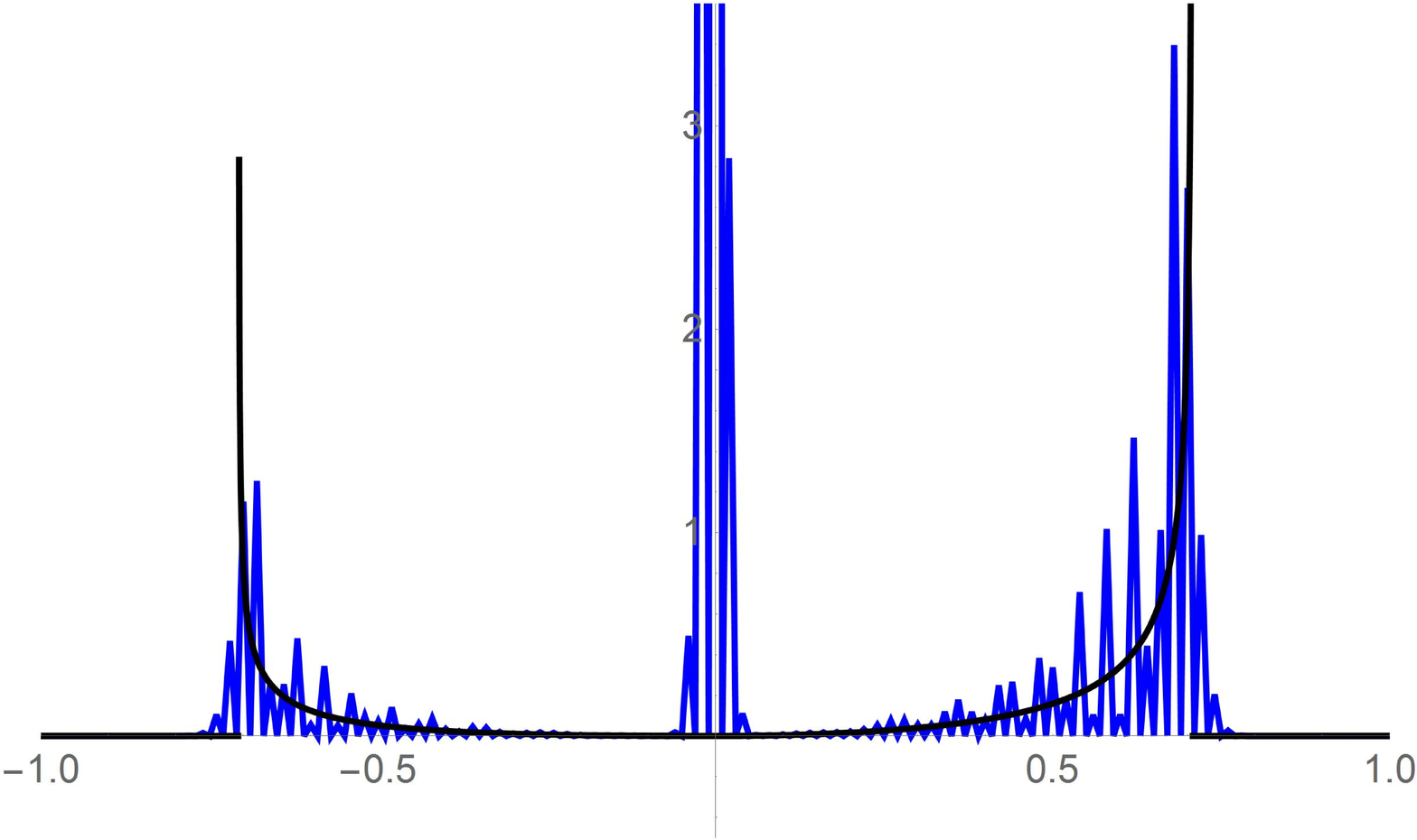, width=7.1cm}} 
\vspace*{13pt}
\caption{\label{fig1.} Case of $\Psi_{0}(0)={}^T\![1,0]$:\\Blue line: Probability Distribution in a re-scaled \\space $(x/100, 100P_{100}(x))$ at time $100$, \\Black line: $w(x)f_{K}(x; 1/\sqrt{2})$}
\end{minipage}
\begin{minipage}{0.5\hsize}
\centerline{\epsfig{file=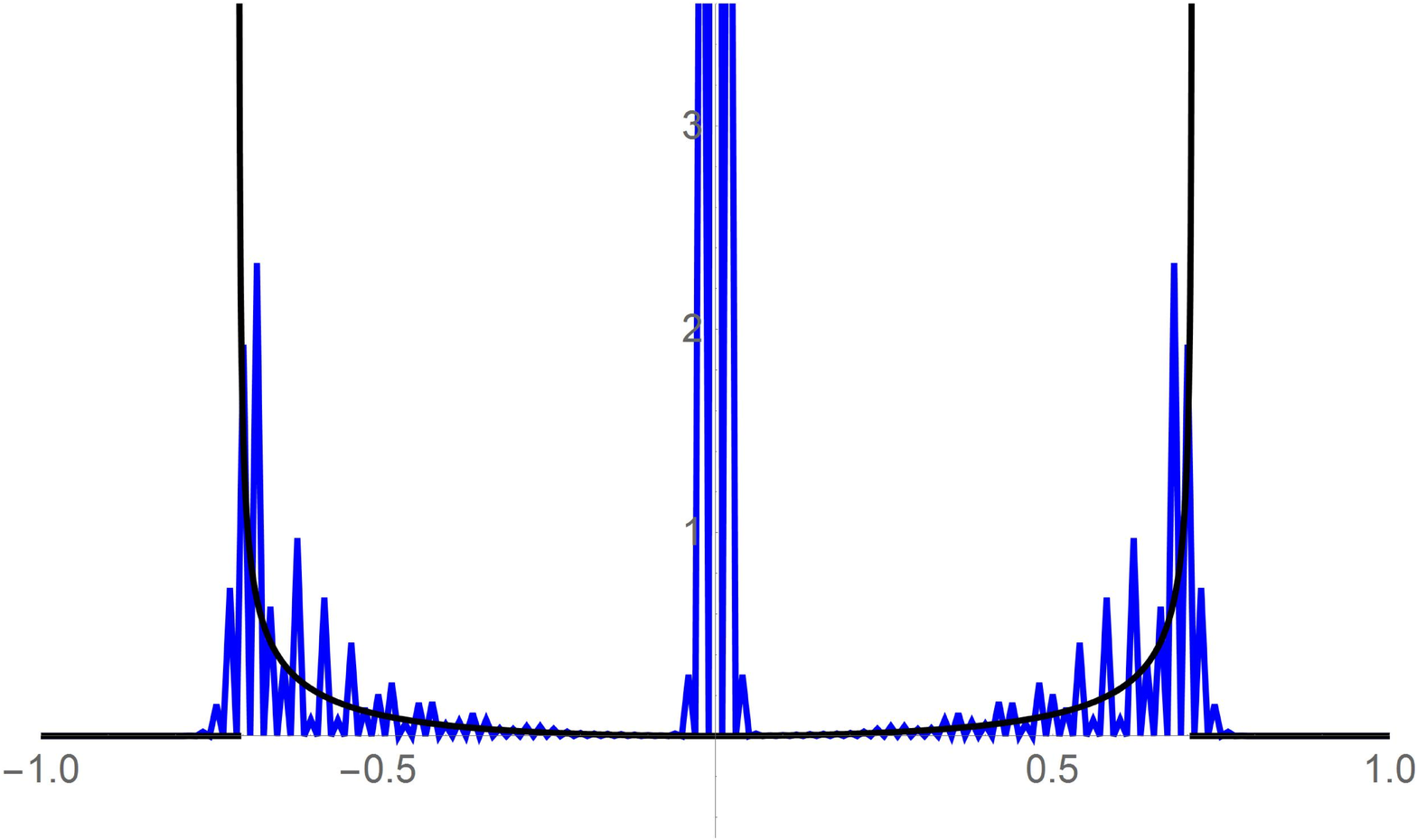, width=7.1cm}} 
\vspace*{13pt}
\caption{\label{fig2.} Case of $\Psi_{0}(0)={}^T\![i/\sqrt{2},1/\sqrt{2}]$:\\Blue line: Probability Distribution in a re-scaled space $(x/100, 100P_{100}(x))$ at time $100$, \\Black line: $w(x)f_{K}(x; 1/\sqrt{2})$}
 \end{minipage}
\end{figure}

\begin{figure}[h]
\begin{minipage}{0.5\hsize}
\centerline{\epsfig{file=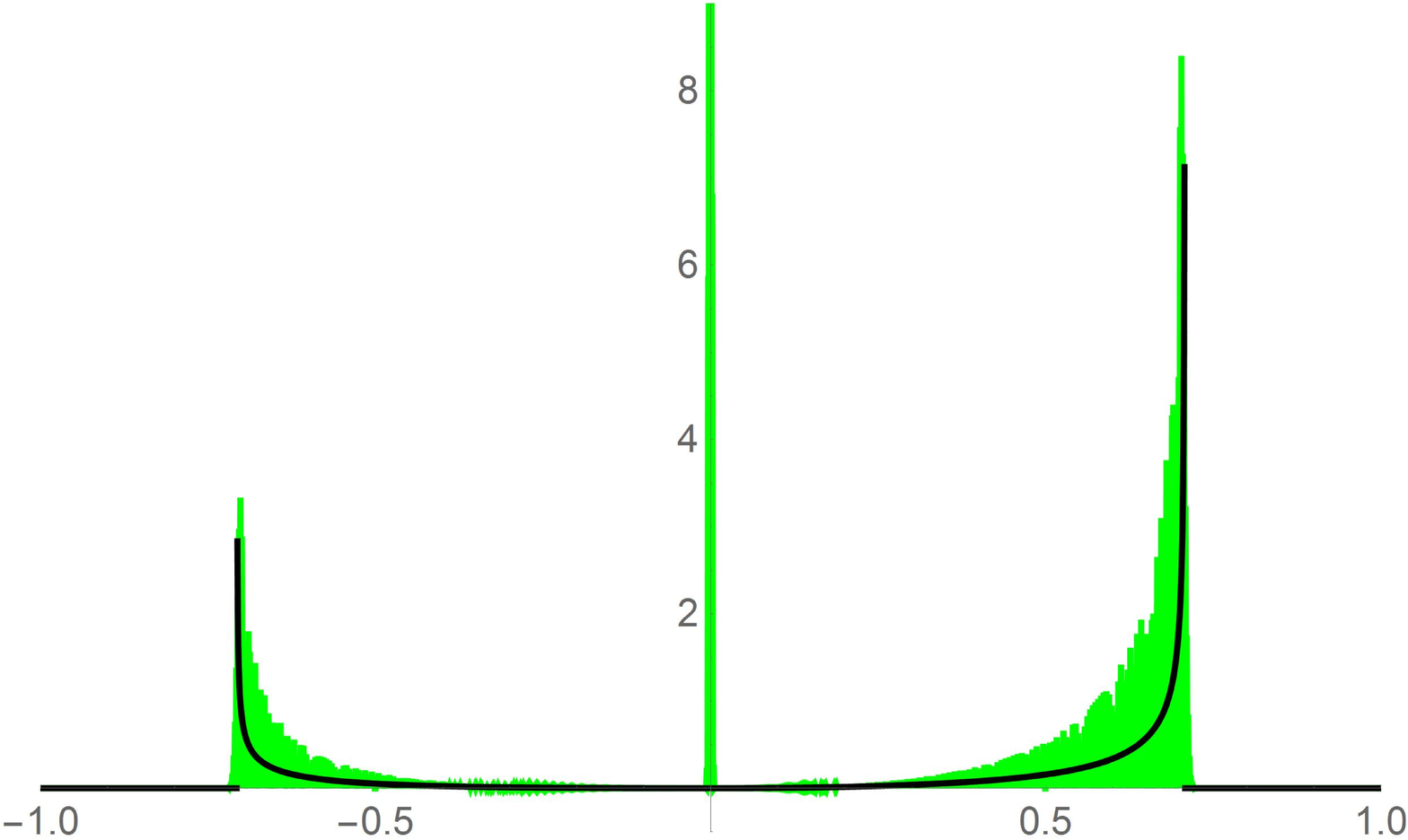, width=7.1cm}} 
\vspace*{13pt}
\caption{\label{fig3.} Case of $\Psi_{0}(0)={}^T\![1,0]$:\\Green line: Probability Distribution in a re-scaled \\space $(x/1000, 1000P_{1000}(x))$ at time $1000$, \\Black line: $w(x)f_{K}(x; 1/\sqrt{2})$}
\end{minipage}
\begin{minipage}{0.5\hsize}
\centerline{\epsfig{file=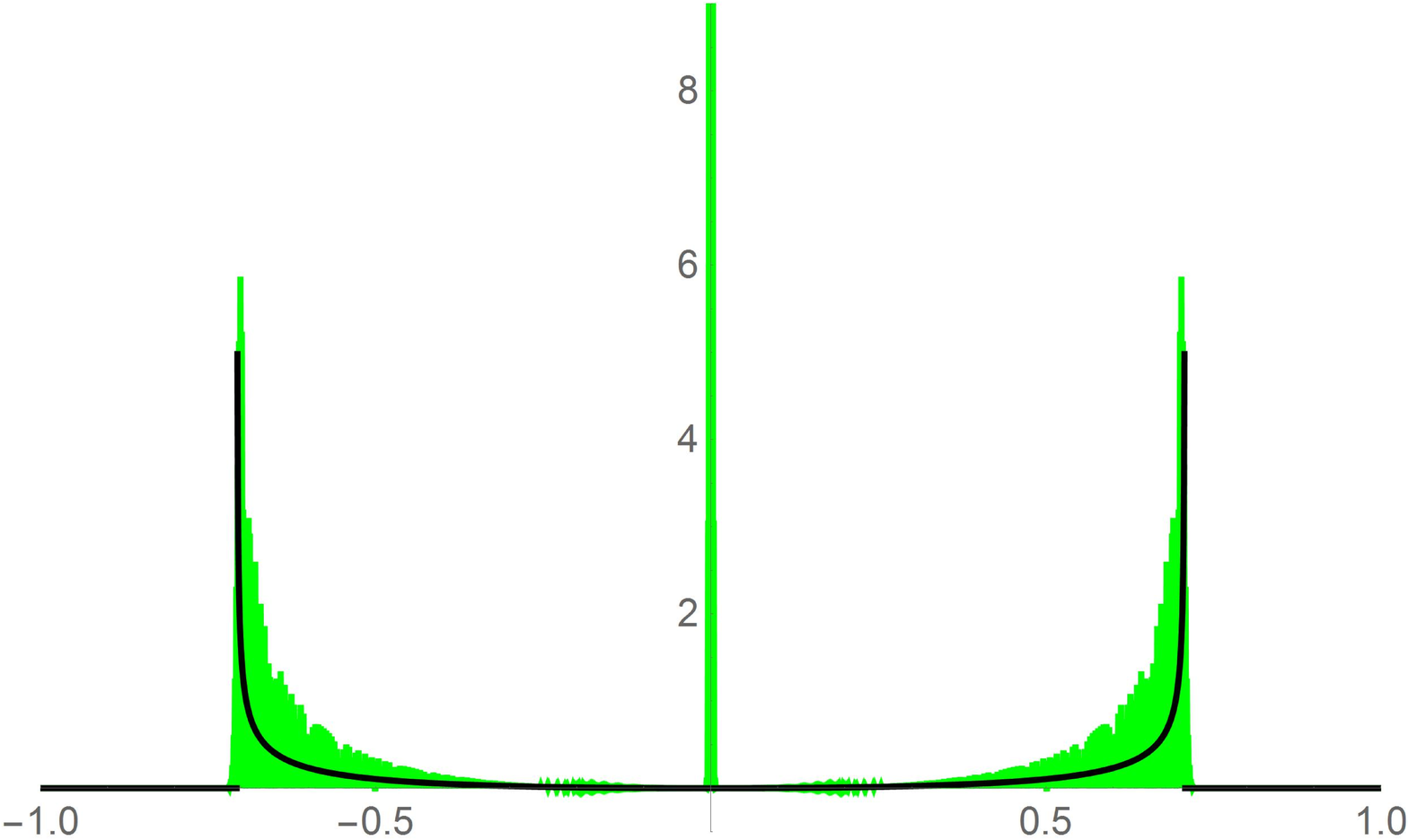, width=7.1cm}} 
\vspace*{13pt}
\caption{\label{fig4.} Case of $\Psi_{0}(0)={}^T\![i/\sqrt{2},1/\sqrt{2}]$:\\Green line: Probability Distribution in a re-scaled space $(x/1000, 1000P_{1000}(x))$ at time $1000$, \\Black line: $w(x)f_{K}(x; 1/\sqrt{2})$}
 \end{minipage}
\end{figure}

\begin{figure}[h]
\begin{minipage}{0.5\hsize}
\centerline{\epsfig{file=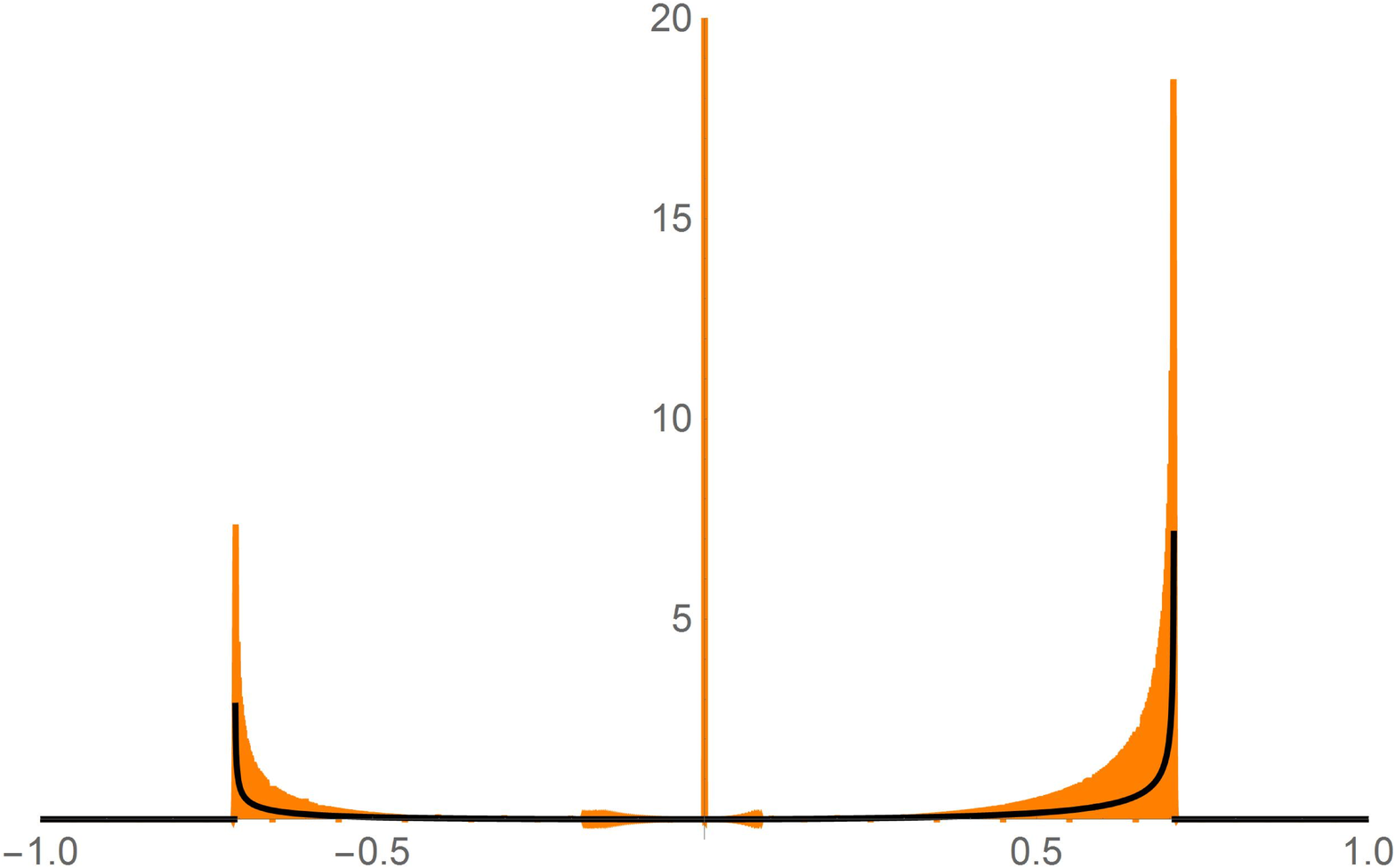, width=7.1cm}} 
\vspace*{13pt}
\caption{\label{fig5.} Case of $\Psi_{0}(0)={}^T\![1,0]$:\\Orange line: Probability Distribution in a re-scaled \\space $(x/10000, 10000P_{10000}(x))$ at time $10000$, \\Black line: $w(x)f_{K}(x; 1/\sqrt{2})$}
\end{minipage}
\begin{minipage}{0.5\hsize}
\centerline{\epsfig{file=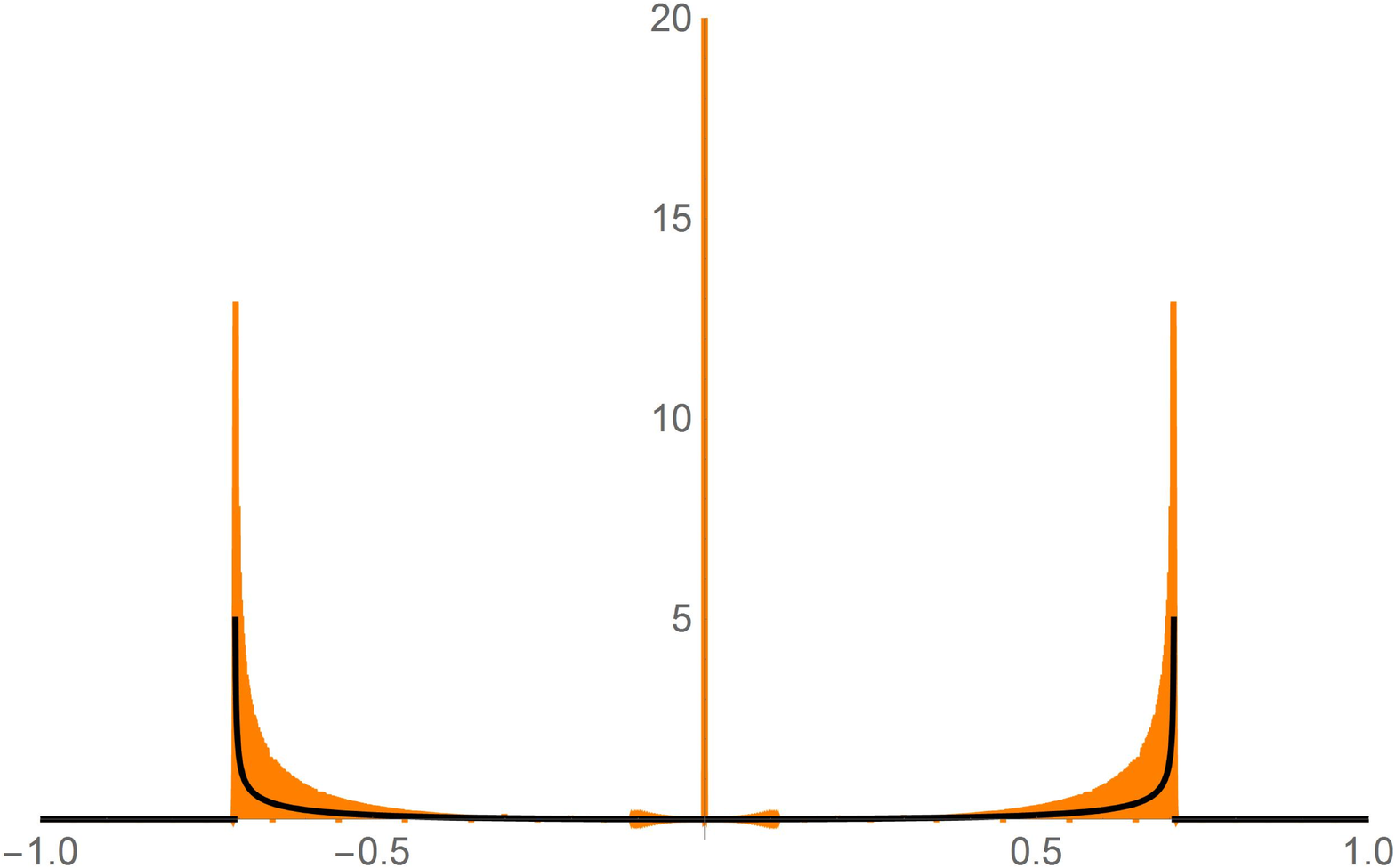, width=7.1cm}} 
\vspace*{13pt}
\caption{\label{fig6.} Case of $\Psi_{0}(0)={}^T\![i/\sqrt{2},1/\sqrt{2}]$:\\Orange line: Probability Distribution in a re-scaled space $(x/10000, 10000P_{10000}(x))$ at time $10000$, \\Black line: $w(x)f_{K}(x; 1/\sqrt{2})$}
 \end{minipage}
\end{figure}

\item Put the initial coin state $\Psi_{0}(0)={}^T\![i/\sqrt{2},1/\sqrt{2}]$.
We obtain from Theorem \ref{weaklimit} the weight fuction $w(x)$ by
\begin{eqnarray}w(x)=\dfrac{3x^{2}}{4+x^{2}}.\end{eqnarray}

From Theorem $2$ in \cite{endo}, we have the coefficient of the delta function $\delta_{0}(x)$ in Eq. \eqref{weakmeasure} by
\begin{align}C=\sum_{x}\overline{\mu}_{\infty}(x)=\dfrac{8}{25}+2\times\dfrac{24}{25}\sum_{y=1}^{\infty}\left(\dfrac{1}{5}\right)^{y}=\dfrac{4}{5},
\end{align}
where
\[
\left\{ \begin{array}{l}
\overline{\mu}_{\infty}(0)=\dfrac{8}{25},\\
\overline{\mu}_{\infty}(x)=\dfrac{24}{25}\left(\dfrac{1}{5}\right)^{|x|}.
\end{array} \right.
\]
\noindent
Accordingly, we have
\begin{align}\int_{-\frac{1}{\sqrt{2}}}^{\frac{1}{\sqrt{2}}}w(x)f_{K}(x;1/\sqrt{2}) dx=\dfrac{1}{5},\end{align}
which suggests
\[C+\int_{-\frac{1}{\sqrt{2}}}^{\frac{1}{\sqrt{2}}}w(x)f_{K}(x;1/\sqrt{2}) dx=1.\]
Now, we show the numerical results of the probability distribution at time $t=100, 1000$, and $10000$ in re-scaled spaces $(x/t, tP_{t}(x))\;(t=100,1000,10000)$. We also give the graph of $w(x)f_{K}(x;1/\sqrt{2})$ on the picture at each time. We see again that the graph of $w(x)f_{K}(x;1/\sqrt{2})$ is just on the middle of the probability distribution for each position at each time, which suggests that our result is appropriate mathematically. 
We also emphasize that $w(x)f_{K}(x;1/\sqrt{2})$ has an origin symmetry (Figs. 2,4,6) , which indicates that the weak limit measure represents the symmetry of the probability distribution (Figs. 2,4,6), and the symmetric initial coin state gives the symmetry. 
\end{enumerate}
\item \underline{{\bf QW with one defect: $\phi=1/4$ case.}}\\
Let us treat the QW whose time evolution is defined by the unitary matrices
\begin{align}
U_{x}=\left\{ \begin{array}{ll}
\dfrac{1}{\sqrt{2}}\begin{bmatrix}1 & 1\\ 1 & -1 \end{bmatrix}\quad (x=\pm1,\pm2,\cdots), \\
\\
\dfrac{i}{\sqrt{2}}\begin{bmatrix}1 & 1\\ 1 & -1 \end{bmatrix} \quad (x=0).\\
\end{array}\right. \label{example4}\end{align}
\par\noindent
The QW is obtained by putting $\phi=1/4$ in Eq. \eqref{wojcik_def}.
\begin{enumerate}
\item 
Put the initial coin state $\Psi_{0}(0)={}^T\![1,0]$.
We get the weight function $w(x)$ in Eq. \eqref{siki} from Theorem \ref{weaklimit} by
\[w(x)=\left\{ \begin{array}{ll}
\dfrac{x^{3}+5x^{2}-2x+2}{x^{2}+4} & (x\geq0), \\
\dfrac{x^{3}-x^{2}-2x+2}{x^{2}+4} & (x<0).
\end{array} \right.\]
Hence, we have
\begin{align}\int_{-\frac{1}{\sqrt{2}}}^{\frac{1}{\sqrt{2}}}w(x)f_{K}(x;1/\sqrt{2}) dx=\dfrac{3}{5}.\end{align}
Now, we should note that we obtained the time-averaged limit measure $\overline{\mu}_{\infty}(x)$ by Theorem $2$ in \cite{endo}, and as a result, we obtain the coefficient of the delta function $\delta_{0}(x)$ in Eq. \eqref{weakmeasure} by
\begin{align}C=\sum_{x}\overline{\mu}_{\infty}(x)=\dfrac{4}{25}+2\times\dfrac{12}{25}\sum_{y=1}^{\infty}\left(\dfrac{1}{5}\right)^{y}=\dfrac{2}{5},
\end{align}
since
\[
\left\{ \begin{array}{l}
\overline{\mu}_{\infty}(0)=\dfrac{4}{25},\\
\overline{\mu}_{\infty}(x)=\dfrac{12}{25}\left(\dfrac{1}{5}\right)^{|x|}.
\end{array} \right.\]
\noindent
Therefore, we see
\[C+\int_{-\frac{1}{\sqrt{2}}}^{\frac{1}{\sqrt{2}}}w(x)f_{K}(x;1/\sqrt{2}) dx=1.\]
Here, we give the numerical results of the probability distribution at time $t=100, 1000$, and $10000$ in re-scaled spaces $(x/t, tP_{t}(x))\;(t=100,1000,10000)$, where $x$ expresses the position of the walker and $P_{t}(x)$ is the probability that the walker exists on position $x$ at time $t$. We should note that $x/t$ corresponds to the real axis, and $tP_{t}(x)$ corresponds to the imaginary axis, respectively. Also, we put the graph of $w(x)f_{K}(x;1/\sqrt{2})$ on the picture at each time. We see that the graph of $w(x)f_{K}(x;1/\sqrt{2})$ is right on the middle of the probability distribution for each position at each time, which suggests that our result is mathematically proper. 
We also note that $\overline{\mu}_{\infty}(x)$ is symmetric for the origin \cite{endo}, however, $w(x)f_{K}(x;1/\sqrt{2})$ does not have an origin symmetry (Figs. 7,9,11) , which suggests that the weak limit measure represents the asymmetry of the probability distribution (Figs. 7,9,11). Moreover, comparing with the results of $\phi=1/2$, we notice that the aymmetry of the probability distributions in the case of $\phi=1/2$ are more prominent than those of $\phi=1/4$.

\begin{figure}[h]
\begin{minipage}{0.5\hsize}
\centerline{\epsfig{file=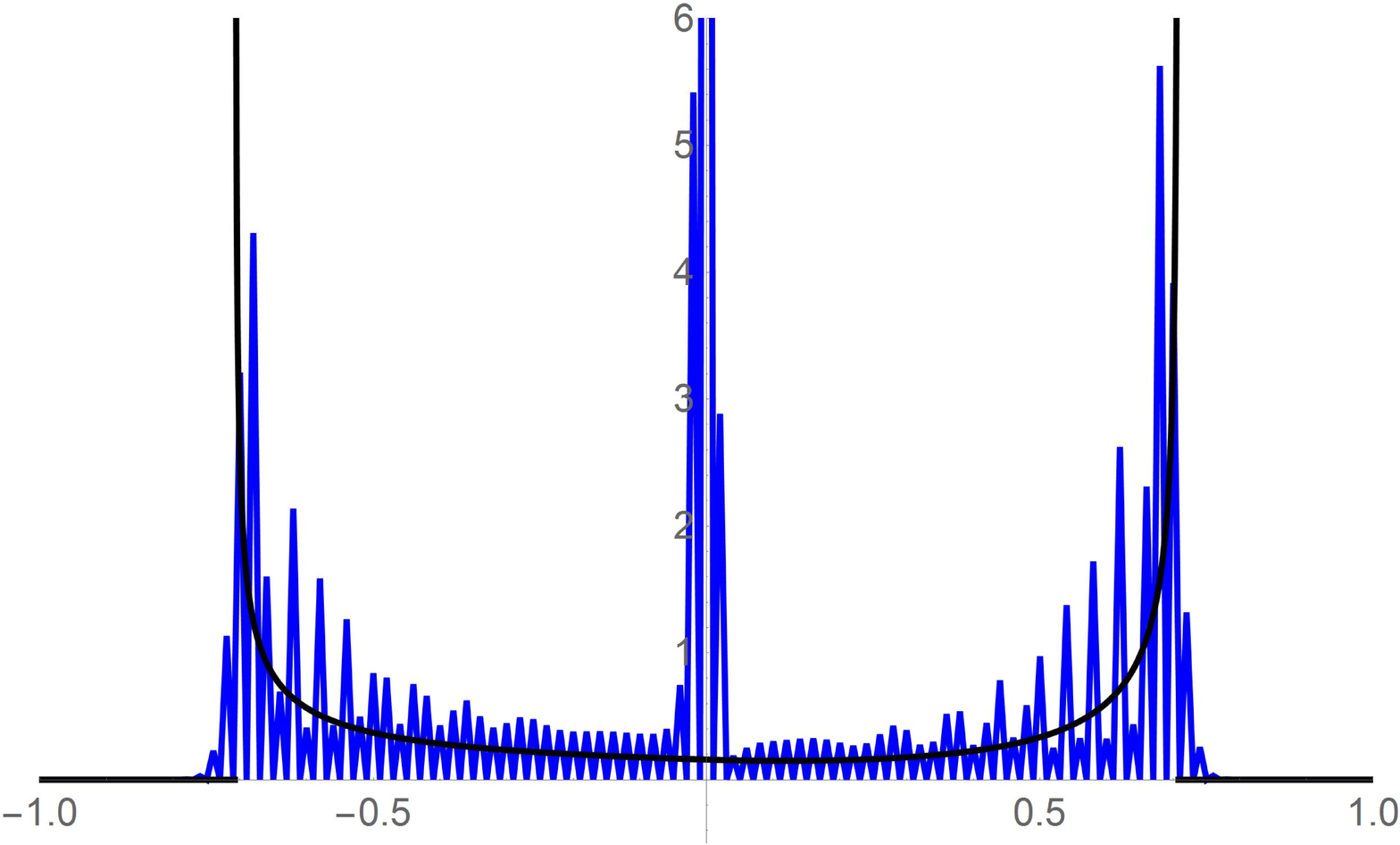, width=7.1cm}} 
\vspace*{13pt}
\caption{\label{fig1.} Case of $\Psi_{0}(0)={}^T\![1,0]$:\\Blue line: Probability Distribution in a re-scaled \\space $(x/100, 100P_{100}(x))$ at time $100$, \\Black line: $w(x)f_{K}(x; 1/\sqrt{2})$}
\end{minipage}
\begin{minipage}{0.5\hsize}
\centerline{\epsfig{file=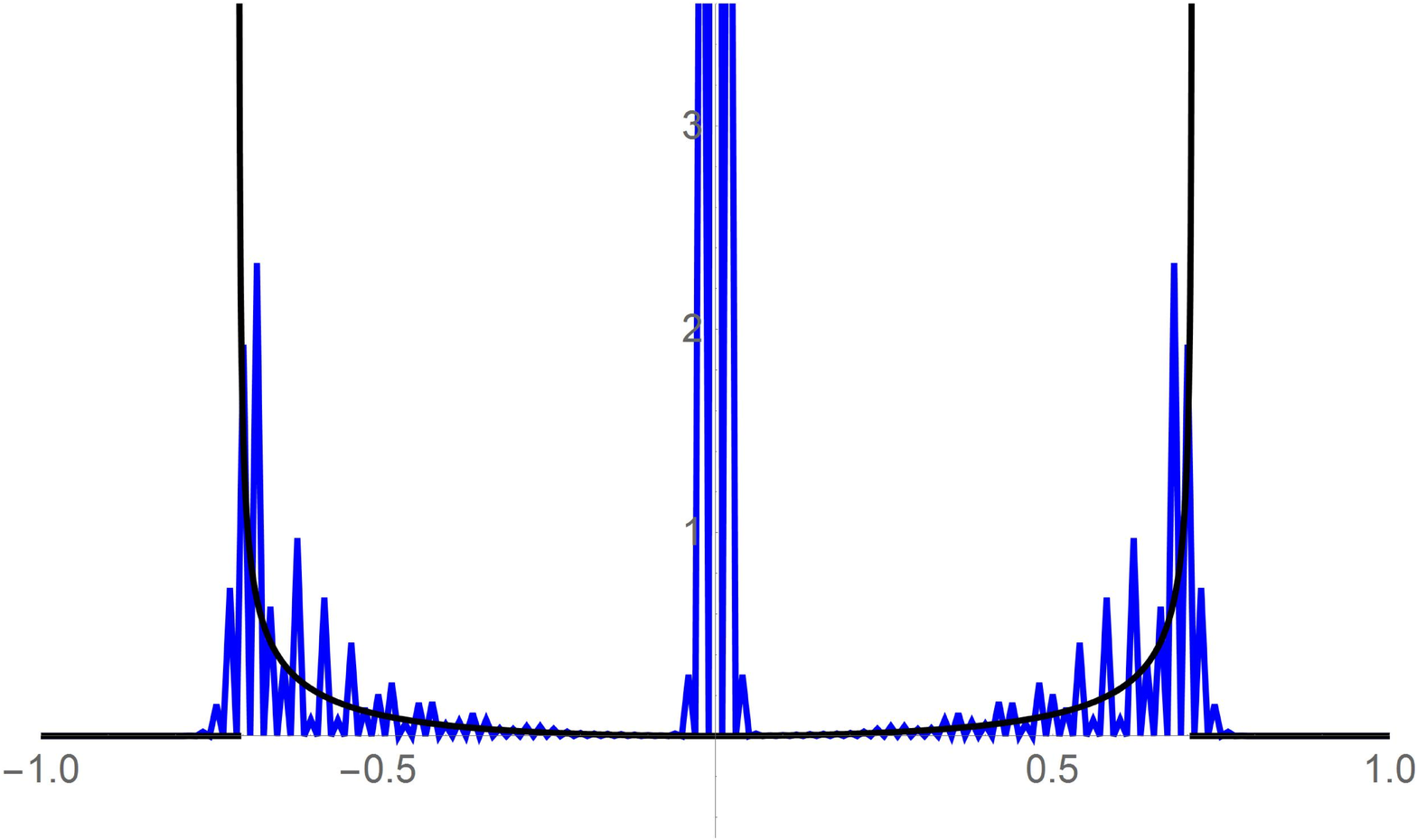, width=7.1cm}} 
\vspace*{13pt}
\caption{\label{fig2.} Case of $\Psi_{0}(0)={}^T\![i/\sqrt{2},1/\sqrt{2}]$:\\Blue line: Probability Distribution in a re-scaled space $(x/100, 100P_{100}(x))$ at time $100$, \\Black line: $w(x)f_{K}(x; 1/\sqrt{2})$}
 \end{minipage}
\end{figure}

\begin{figure}[h]
\begin{minipage}{0.5\hsize}
\centerline{\epsfig{file=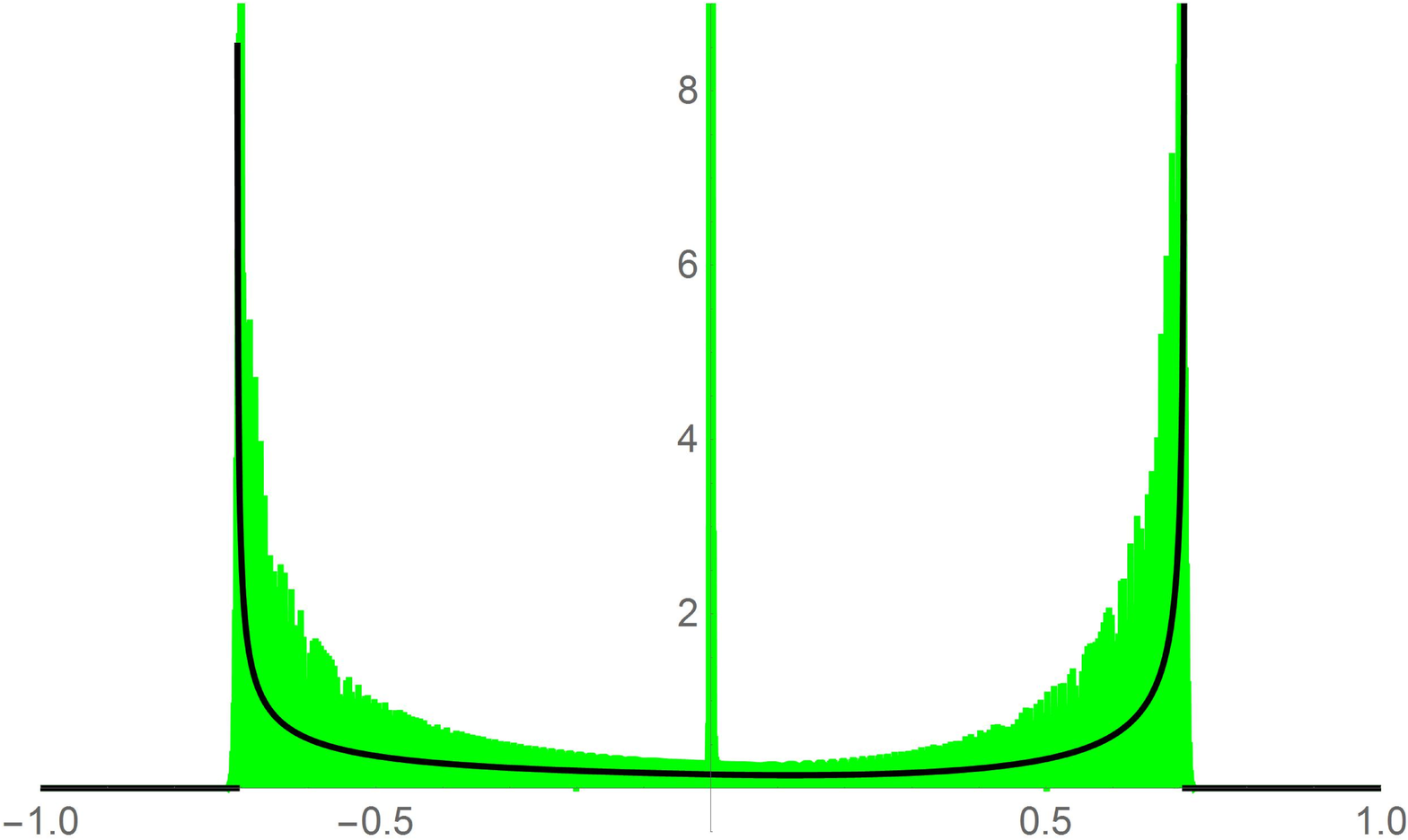, width=7.1cm}} 
\vspace*{13pt}
\caption{\label{fig3.} Case of $\Psi_{0}(0)={}^T\![1,0]$:\\Green line: Probability Distribution in a re-scaled \\space $(x/1000, 1000P_{1000}(x))$ at time $1000$, \\Black line: $w(x)f_{K}(x; 1/\sqrt{2})$}
\end{minipage}
\begin{minipage}{0.5\hsize}
\centerline{\epsfig{file=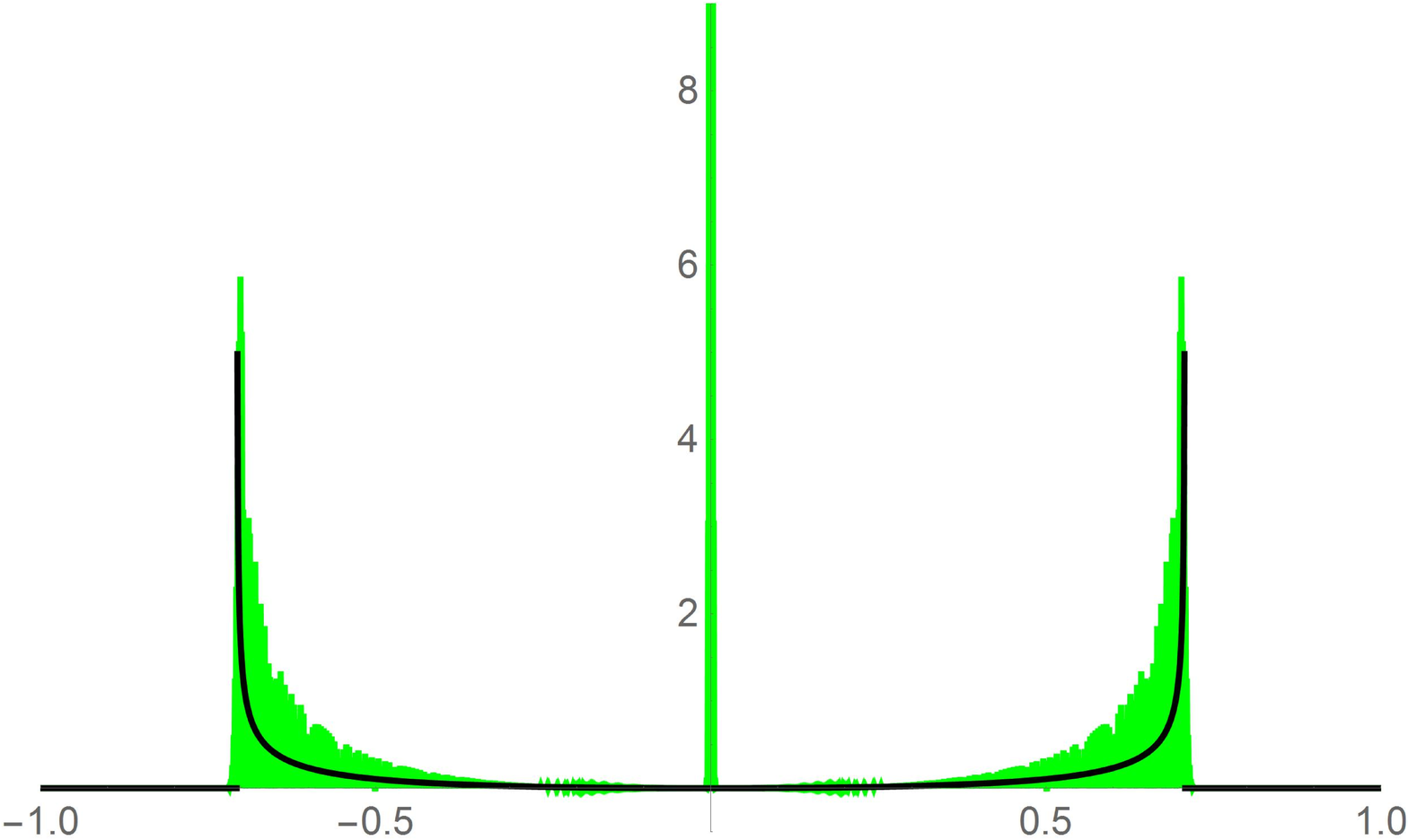, width=7.1cm}} 
\vspace*{13pt}
\caption{\label{fig4.} Case of $\Psi_{0}(0)={}^T\![i/\sqrt{2},1/\sqrt{2}]$:\\Green line: Probability Distribution in a re-scaled space $(x/1000, 1000P_{1000}(x))$ at time $1000$, \\Black line: $w(x)f_{K}(x; 1/\sqrt{2})$}
 \end{minipage}
\end{figure}

\begin{figure}[h]
\begin{minipage}{0.5\hsize}
\centerline{\epsfig{file=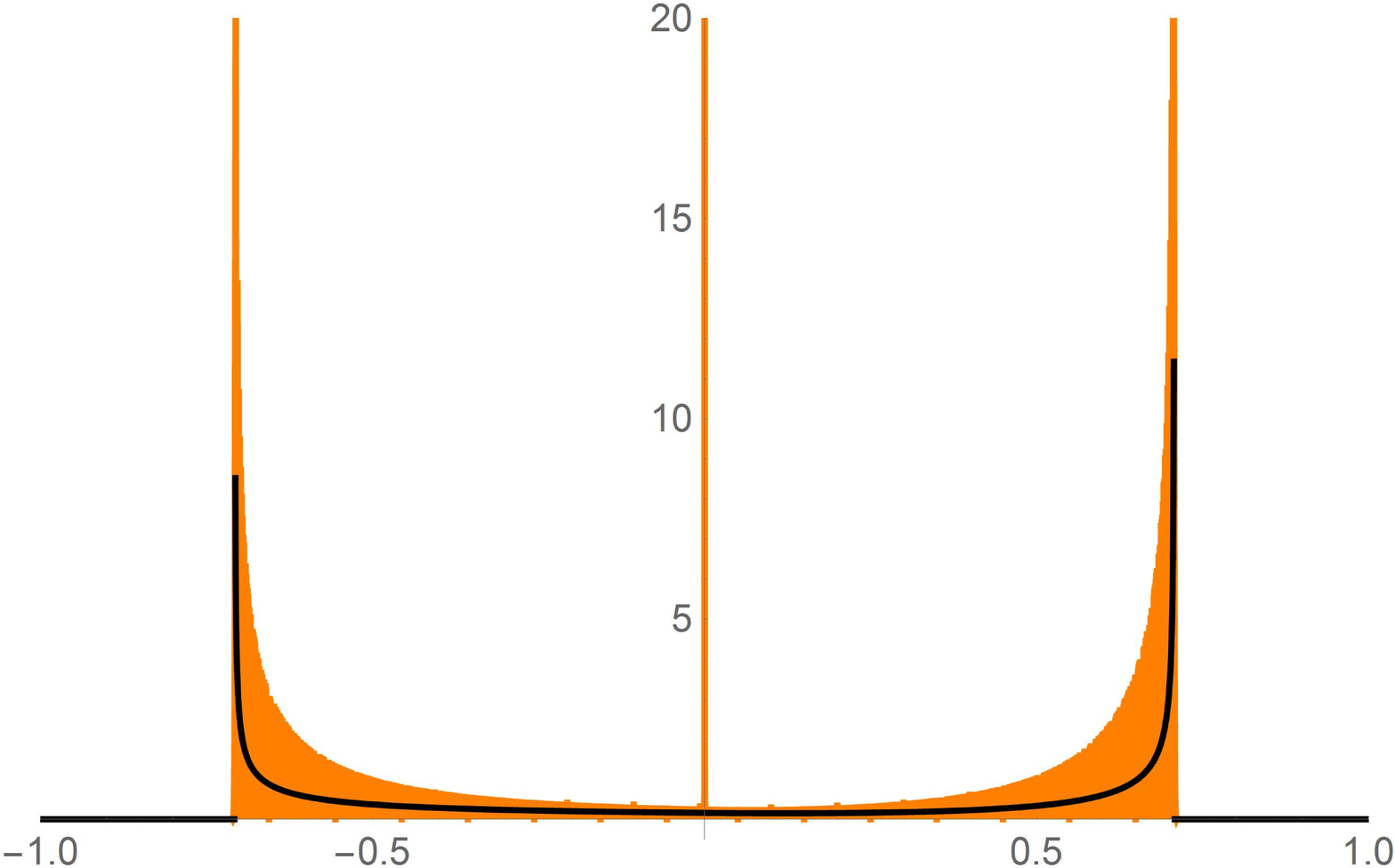, width=7.1cm}} 
\vspace*{13pt}
\caption{\label{fig5.} Case of $\Psi_{0}(0)={}^T\![1,0]$:\\Orange line: Probability Distribution in a re-scaled \\space $(x/10000, 10000P_{10000}(x))$ at time $10000$, \\Black line: $w(x)f_{K}(x; 1/\sqrt{2})$}
\end{minipage}
\begin{minipage}{0.5\hsize}
\centerline{\epsfig{file=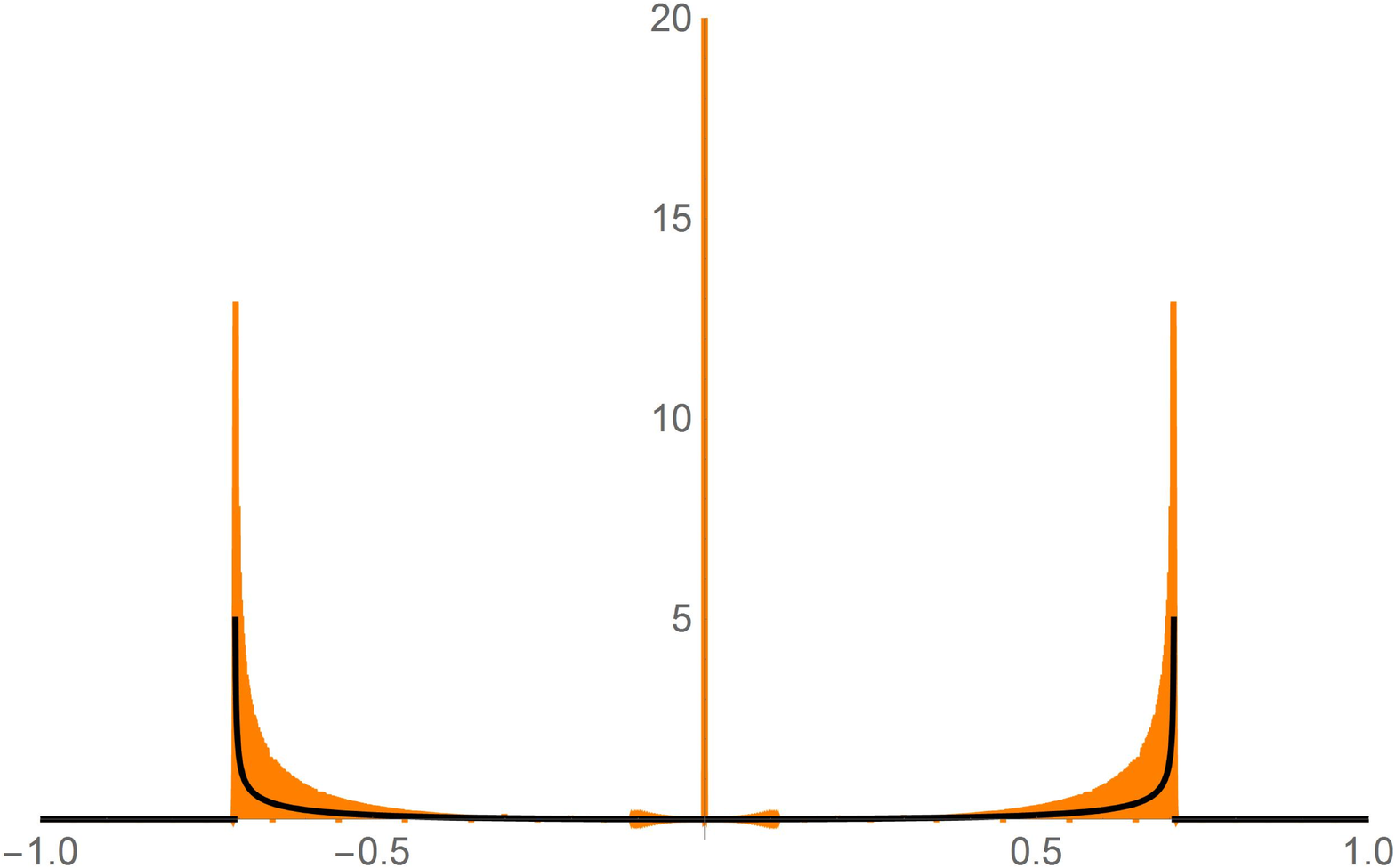, width=7.1cm}} 
\vspace*{13pt}
\caption{\label{fig6.} Case of $\Psi_{0}(0)={}^T\![i/\sqrt{2},1/\sqrt{2}]$:\\Orange line: Probability Distribution in a re-scaled space $(x/10000, 10000P_{10000}(x))$ at time $10000$, \\Black line: $w(x)f_{K}(x; 1/\sqrt{2})$}
 \end{minipage}
\end{figure}

\noindent
\item Let the initial coin state be $\Psi_{0}(0)={}^T\![i/\sqrt{2},1/\sqrt{2}]$.
Then the weight fuction $w(x)$ is given from Theorem \ref{weaklimit} by
\begin{eqnarray}w(x)=\dfrac{3x^{2}}{4+x^{2}}.\end{eqnarray}
\noindent
From Theorem $2$ in \cite{endo}, we have the coefficient of the delta function $\delta_{0}(x)$ in Eq. \eqref{weakmeasure} by
\begin{align}C=\sum_{x}\overline{\mu}_{\infty}(x)=\dfrac{8}{25}+2\times\dfrac{24}{25}\sum_{y=1}^{\infty}\left(\dfrac{1}{5}\right)^{y}=\dfrac{4}{5},
\end{align}
where
\[
\left\{ \begin{array}{l}
\overline{\mu}_{\infty}(0)=\dfrac{8}{25},\\
\overline{\mu}_{\infty}(x)=\dfrac{24}{25}\left(\dfrac{1}{5}\right)^{|x|}.
\end{array} \right.
\]
\noindent
Thereby, we have
\begin{align}\int_{-\frac{1}{\sqrt{2}}}^{\frac{1}{\sqrt{2}}}w(x)f_{K}(x;1/\sqrt{2}) dx=\dfrac{1}{5},\end{align}
which suggests
\[C+\int_{-\frac{1}{\sqrt{2}}}^{\frac{1}{\sqrt{2}}}w(x)f_{K}(x;1/\sqrt{2}) dx=1.\]
Now, we show the numerical results of the probability distribution at time $t=100, 1000$, and $10000$ in re-scaled spaces $(x/t, tP_{t}(x))\;(t=100,1000,10000)$. We also give the graph of $w(x)f_{K}(x;1/\sqrt{2})$ on the picture at each time. We see again that the graph of $w(x)f_{K}(x;1/\sqrt{2})$ is just on the middle of the probability distribution for each position at each time, which suggests that our result is appropriate mathematically. 
We also emphasize that $w(x)f_{K}(x;1/\sqrt{2})$ is symmetric for the origin (Figs. 8,10,12) , which indicates that the weak limit measure represents the symmetry of the probability distribution (Figs. 8,10,12), and the symmetric initial coin state contributes to the symmetry.

\end{enumerate}
\end{enumerate}

\section{Summary}

In this paper, we obtained the weak convergence theorem for the Wojcik model, and give numerical results of the probability distribution for some conclete phase parameters of the defect and initial coin states.
The purpose of this work is to clarify mathematically the whole picture of the asymptotic behavior of the Wojcik model. Especially, one of the motivations is to investigate the effect of the phase and initial coin state of the walker on the probability distribution. 
Our analytical result, that is the weight function in the weak convergence theorem suggests in general the asymmetry of the probability distribution.
Throughout our study including the numerical results, it can be expected that both the phase and initial coin state heavily influence on the aymptotic behavior of the walker.
Indeed, as we saw in Section \ref{examples}, the symmetry of the initial coin states give the symmetric distributions, and the asymmetric initial coin states contribute to the asymmetric distributions.
Moreover, if the phase gradually approximates to 0, then the probability distribution seems also close to that of the Hadamard walk \cite{kempesan}.\vspace{5mm}\\

\noindent
{\bf Acknowledgments.}
We would like to thank Shimpei Endo for giving useful advices to the numerical simulations.
NK acknowledges financial support of the Grant-in-Aid for Scientific Research (C) of Japan Society for the
Promotion of Science (Grant No. 24540116).
\vspace{5mm}
\begin{small}
\bibliographystyle{jplain}

\end{small}
\vspace{5mm}
\noindent {\large{\bf Appendix A}}  \\
\noindent
In Appendix A, we give the proof of Theorem \ref{weaklimit}. 
Throughout the proof, we focus on the characteristic function of the Wojcik model in the long-time limit, that is,
\begin{align}E\left[e^{i\xi \frac{X_{t}}{t}}\right]=\int_{x\in\mathbb{Z}}g_{X_{t}/t}(x)e^{i\xi x}dx\quad(t\to\infty)\label{density_function}\end{align}
where $g_{X_{t}/t}(x)$ is the density function of $X_{t}/t$. Hereafter, we derive $E\left[e^{i\xi X_{t}/t}\right]$ in the long-time limit.
We should remark that to get the second term of Eq. \eqref{siki}, $w(x)f_{K}(x;1/\sqrt{2})$, corresponds to obtain $g_{X_{t}/t}(x)\;(t\to\infty)$.\\
Now, we introduce the weight of all the paths of the walker, which moves left $l$ times and moves right $m$ times till time $t$ \cite{segawa};
\[\Xi_{t}(x)=\sum_{l_{j},m_{j}}P_{x_{l_{1}}}^{l_{1}}Q_{x_{m_{1}}}^{m_{1}}P_{x_{l_{2}}}^{l_{2}}Q_{x_{m_{2}}}^{m_{2}}\cdots P_{x_{l_{t}}}^{l_{t}}Q_{x_{m_{t}}}^{m_{t}},\]\\
where $l+m=t,\;-l+m=x,\;\;\sum_{i}l_{i}=l,\;\sum_{j}m_{j}=m$ with $l_{i}+m_{i}=1,\;l_{i},m_{i}\in\{0,1\}$, and $\;\sum_{\gamma=l_{i},m_{j}}|x_{\gamma}|=x$. 
Here, we consider $z\in\mathbb{C}$ on a unit circle.
According to \cite{segawa}, $E[e^{i\xi X_{t}/t}]\;(t\to\infty)$ is expressed by 
square norm of the residue of $\tilde{\Xi}_{x}(z)=\sum_{t}\Xi_{t}(x)z^{t}$ as 
\begin{align}
E\left[e^{i\xi\frac{X_{t}}{t}}\right]\to\int^{2\pi}_{0}\sum_{\theta\in B}e^{-i\xi \theta^{'}(k)}\|Res(\hat{\tilde{\Xi}}(k:z):z=e^{i\theta(k)})\|^{2}\dfrac{dk}{2\pi}\qquad(t\to\infty),\label{limitdensity}
\end{align}
where $B$ is the set of the singular points of $\hat{\tilde{\Xi}}(k:z)\equiv\sum_{x\in\mathbb{Z}}\tilde{\Xi}_{x}(z)e^{ikx}$. Note $\theta^{'}(k)=d \theta(k)/d k$. 
We will explain how to derive Eq. \eqref{limitdensity} in Appendix A. 
Using Eq. \eqref{limitdensity} mainly, we prove Theorem \ref{weaklimit}.
Now, we set worthwhile expressions of $\tilde{\Xi}_{x}(z)$ which play important roles in the proof. 
Lemma \ref{prf2.2} corresponds to Lemma $5$ in \cite{endo}, which we also took advantage of to derive the time-averaged limit measure of the Wojcik model. 
Let the quantum walker start from the origin with the initial coin state $\Psi_{0}(0)={}^T\![\alpha,\beta]$ with $\alpha,\beta\in\mathbb{C}$
and $|\alpha|^{2}+|\beta|^{2}=1$.
\begin{lemma}
\label{prf2.2}
\begin{enumerate}
\item If $x=0$, we have 
\[\tilde{\Xi}_{0}(z)=\dfrac{1}{1-\sqrt{2}\omega\tilde{f}_{0}(z)+\omega^{2}\{\tilde{f}_{0}(z)\}^{2}}
\begin{bmatrix} 
1-\dfrac{e^{2\pi i\phi}}{\sqrt{2}}\tilde{f}_{0}(z) & -\dfrac{e^{2\pi i\phi}}{\sqrt{2}}\tilde{f}_{0}(z)\\
&\\
\dfrac{e^{2\pi i\phi}}{\sqrt{2}}\tilde{f}_{0}(z) & 1-\dfrac{e^{2\pi i\phi}}{\sqrt{2}}\tilde{f}_{0}(z)\\
\end{bmatrix}.\]

\item If $|x|\geq 1$, we have
\[
\tilde{\Xi}_{x}(z)=\left\{\begin{array}{ll}
(\tilde{\lambda}_{x}(z))^{x-1}
\left[
\begin{array}{c}
\tilde{\lambda}_{x}(z)\tilde{f}_{x}(z)\\
z \\
\end{array}
\right]\left[\dfrac{e^{2\pi i\phi}}{\sqrt{2}},-\dfrac{e^{2\pi i\phi}}{\sqrt{2}}\right]\tilde{\Xi}_{0}(z) & (x\geq 1), \\
&\\
(\tilde{\lambda}_{x}(z))^{|x|-1}
\left[
\begin{array}{c}
z \\
\tilde{\lambda}_{x}(z)\tilde{f}_{x}(z) \\
\end{array}
\right]\left[\dfrac{e^{2\pi i\phi}}{\sqrt{2}},\dfrac{e^{2\pi i\phi}}{\sqrt{2}}\right]\tilde{\Xi}_{0}(z) & (x\leq -1), \\
\end{array} \right.\]
\end{enumerate}
where $\tilde{\lambda}_{x}(z)=\dfrac{z}{\tilde{f}_{x}(z)-\sqrt{2}}$. 
Here $\tilde{f}_{x}(z)$ satisfies the following quadratic equation.
\begin{align}(\tilde{f}_{x}(z))^{2}-\sqrt{2}(1+z^{2})\tilde{f}_{x}(z)+z^{2}=0.\label{(Takako1)}\end{align}
\end{lemma}
\noindent
Hereafter, we write $\tilde{f}_{x}(z)$ by $\tilde{f}(z)$, since $\tilde{f}^{(\pm)}_{x}(z)$ do not depend on the position. Note that 
$\tilde{f}_{x}(z)$ is originated from $\tilde{f}^{(\pm)}_{x}(z)$ which satisfy \cite{segawa}
\begin{align*}
\tilde{f}^{(\pm)}_{x}(z)=\sqrt{2}z^{2}\left\{1-\dfrac{1}{2-\sqrt{2} \tilde{f}^{(\pm)}_{x\pm1}(z)}\right\}.
\vspace{5mm}\end{align*}
First, we derive the singular points of $\hat{\tilde{\Xi}}(k:z)$ and then, calculate the residues of $\hat{\tilde{\Xi}}(k:z)$ at the singular points.
By means of Lemma $1$, we can write down $\hat{\tilde{\Xi}}(k:z)$ by
\begin{eqnarray}\hat{\tilde{\Xi}}(k:z)=\dfrac{e^{ik}}{1-e^{ik}\tilde{\lambda}^{(+)}(z)}\begin{bmatrix}\tilde{\lambda}^{(+)}(z)\tilde{f}(z)\\ z\end{bmatrix}\left[\dfrac{e^{2\pi i\phi}}{\sqrt{2}},-\dfrac{e^{2\pi i\phi}}{\sqrt{2}}\right]\tilde{\Xi}_{0}(z)\nonumber\\
+\dfrac{e^{-ik}}{1-e^{-ik}\tilde{\lambda}^{(-)}(z)}\begin{bmatrix}z\\ \tilde{\lambda}^{(-)}(z)\tilde{f}(z)\end{bmatrix}\left[\dfrac{e^{2\pi i\phi}}{\sqrt{2}},\dfrac{e^{2\pi i\phi}}{\sqrt{2}}\right]\tilde{\Xi}_{0}(z)+\tilde{\Xi}_{0}(z).\label{french}\end{eqnarray}
\noindent
The first term corresponds to the positive part of $\tilde{\Xi}_{x}(z)$, and the second term corresponds to the negative part of $\tilde{\Xi}_{x}(z)$, respectively. We should also note that the singular points derived from the third term, $\tilde{\Xi}_{0}(z)$, contributes to localization.
\noindent
Note that if $|z|<1$, then $|\tilde{\lambda}^{(\pm)}(z)|<1$ should hold. Hence, the infinite series $\sum_{x}(\tilde{\lambda}^{(+)}(z))^{x-1}e^{ikx}$ and $\sum_{x}(\tilde{\lambda}^{(-)}(z))^{|x|-1}e^{-ikx}$ converge.
Here, we have 
\begin{align}
\left\{
\begin{array}{l}
\tilde{\lambda}^{(\pm)}(e^{i\theta})=\mp\{\operatorname{sgn}(\cos\theta)\sqrt{2\cos^{2}\theta-1}+i\sqrt{2}\sin\theta\},\\
\\
\tilde{f}(e^{i\theta})=\operatorname{sgn}(\cos\theta)e^{i\theta}\{\sqrt{2}|\cos\theta|-\sqrt{2\cos^{2}\theta-1}\},
\end{array}
\right.\label{nippon}
\end{align}
\noindent
which will be explained how to derive in Appendix B.
Now, the principal singular points in this paper come from the denominators of the first and second terms of Eq. \eqref{french}, that is,
\begin{eqnarray}1-e^{ik}\tilde{\lambda}^{(+)}(z)=0,\label{eqt.1}\end{eqnarray}and
\begin{eqnarray}1-e^{-ik}\tilde{\lambda}^{(-)}(z)=0.\label{eqt.2}\end{eqnarray}
\noindent
Equations \eqref{eqt.1} and \eqref{eqt.2} give the two conditions for the solutions.
For Eq. \eqref{eqt.1}, we see 
\begin{eqnarray}\cos k=-\operatorname{sgn}(\cos\theta^{(+)}(k))\sqrt{2\cos^{2}\theta^{(+)}(k)-1},\label{cosk+}\end{eqnarray}
\begin{eqnarray}\sin k=\sqrt{2}\sin\theta^{(+)}(k),\label{sink+}\end{eqnarray}
and for Eq. \eqref{eqt.2}, we have
\begin{eqnarray}\cos k=\operatorname{sgn}(\cos\theta^{(-)}(k))\sqrt{2\cos^{2}\theta^{(-)}(k)-1},\label{cosk-}\end{eqnarray}
\begin{eqnarray}\sin k=\sqrt{2}\sin\theta^{(-)}(k).\label{sink-}\end{eqnarray}
\noindent
Here, we put $-d\theta^{(\pm)}(k)/d k=x_{\pm}$ to calculate the RHS of Eq. \eqref{limitdensity} and derive the density function $g_{X_{t}/t}(x)\;\;\;(t\to\infty)$.
Then, we derivate Eqs. \eqref{cosk+} and \eqref{cosk-} for $k$, and we obtain $\sin k,\;\cos k,\;\sin\theta^{(\pm)}(k)$, and $\cos\theta^{(\pm)}(k)$ as follows;
Eqs. \eqref{cosk+} and \eqref{sink+} give
\begin{eqnarray}
\left\{
\begin{array}{l}
\cos k=\operatorname{sgn}(\cos k)\dfrac{x_{+}}{\sqrt{1-x_{+}^{2}}},\;\cos\theta^{(+)}(k)=-\operatorname{sgn}(\cos k)\dfrac{1}{\sqrt{2(1-x_{+}^{2})}},\\
\\
\sin k=\operatorname{sgn}(\sin k)\sqrt{\dfrac{1-2x_{+}^{2}}{1-x_{+}^{2}}},\;\;\sin\theta^{(+)}(k)=\operatorname{sgn}(\sin k)\sqrt{\dfrac{1-2x_{+}^{2}}{2(1-x_{+}^{2})}}.
\end{array}
\right.\label{solutions+}
\end{eqnarray} 
\noindent
Eqs. (\ref{cosk-}) and (\ref{sink-}) yield
\begin{eqnarray}
\left\{
\begin{array}{l}
\cos k=-\operatorname{sgn}(\cos k)\dfrac{x_{-}}{\sqrt{1-x_{-}^{2}}},\;\cos\theta^{(-)}(k)=\operatorname{sgn}(\cos k)\dfrac{1}{\sqrt{2(1-x_{-}^{2})}},\\
\\
\sin k=\operatorname{sgn}(\sin k)\sqrt{\dfrac{1-2x_{-}^{2}}{1-x_{-}^{2}}},\;\;\sin\theta^{(-)}(k)=\operatorname{sgn}(\sin k)\sqrt{\dfrac{1-2x_{-}^{2}}{2(1-x_{-}^{2})}}.
\end{array}
\right.\label{solutions-}
\end{eqnarray} 
\noindent
Thus, we obtain the set of the singular points of $\hat{\tilde{\Xi}}(k:z)$;
\[B=\{e^{i\theta^{(+)}(k)},e^{i\theta^{(-)}(k)}\},\]
where
\[e^{i\theta^{(+)}(k)}=-\dfrac{\operatorname{sgn}(\cos k)}{\sqrt{2(1-x^{2}_{+})}}+i\operatorname{sgn}(\sin k)\sqrt{\dfrac{1-2x^{2}_{+}}{2(1-x^{2}_{+})}},\]
and 
\[e^{i\theta^{(-)}(k)}=\dfrac{\operatorname{sgn}(\cos k)}{\sqrt{2(1-x^{2}_{-})}}+i\operatorname{sgn}(\sin k)\sqrt{\dfrac{1-2x^{2}_{-}}{2(1-x^{2}_{-})}}.\]
Next, we derive the residue of $\hat{\tilde{\Xi}}(k;z)$ at $e^{i\theta^{(\pm)}(k)}$.
Substituting the singular points to $\tilde{f}(z)$, we have
\[\tilde{f}(e^{i\theta^{(+)}(k)})=-\operatorname{sgn}(\cos k)e^{i\theta^{(+)}(k)}\dfrac{\sqrt{1-x^{2}}}{1+|x|},\;\;\tilde{f}(e^{i\theta^{(-)}(k)})=\operatorname{sgn}(\cos k)e^{i\theta^{(-)}(k)}\dfrac{\sqrt{1-x^{2}}}{1+|x|}.\]
\noindent
Owing to Lemma \ref{prf2.2}, we see
\begin{eqnarray*}\!\!\!&\dfrac{e^{ik}}{1-e^{ik}\tilde{\lambda}^{(+)}(z)}&\!\!\!\begin{bmatrix}\tilde{f}(z)\tilde{\lambda}^{(+)}(z)\\ z\end{bmatrix}\left[\dfrac{e^{2\pi i \phi}}{\sqrt{2}},-\dfrac{e^{2\pi i \phi}}{\sqrt{2}}\right]\tilde{\Xi}_{0}(z)\\
\hspace{-10mm}&=&\hspace{-10mm}\dfrac{e^{2\pi i \phi}}{\sqrt{2}}\dfrac{1}{\tilde{\Lambda}_{0}(z)}\dfrac{e^{ik}}{1-e^{ik}\tilde{\lambda}^{(+)}(z)}\begin{bmatrix}\tilde{f}(z)\tilde{\lambda}^{(+)}(z)\\ z\end{bmatrix}\{\alpha-\beta-\sqrt{2}\omega\alpha\tilde{f}(z)\}.\end{eqnarray*}
\par\noindent
By definition, the square norm of residue of the first term of Eq. (\ref{french}) is written by\\

$\left|Res\left(\dfrac{e^{ik}}{1-e^{ik}\tilde{\lambda}^{(+)}(z)}\begin{bmatrix}\tilde{f}(z)\tilde{\lambda}^{(+)}(z)\\ z\end{bmatrix}\left[\dfrac{e^{2\pi i \phi}}{\sqrt{2}},-\dfrac{e^{2\pi i \phi}}{\sqrt{2}}\right]\tilde{\Xi}_{0}(z):z=e^{i\theta^{(+)}(k)}\right)\right|^{2}$
\begin{eqnarray*}&=&\!\!\!\!\dfrac{1}{2}\left|Res\left(\dfrac{1}{1-e^{ik}\tilde{\lambda}^{(+)}(z)}:z=e^{i\theta^{(+)}(k)}\right)\right|^{2}\dfrac{1}{|\tilde{\Lambda}_{0}(e^{i\theta^{(+)}(k)})|^{2}}\left|\begin{bmatrix}\tilde{f}(e^{i\theta^{(+)}(k)})\tilde{\lambda}^{(+)}(e^{i\theta^{(+)}(k)})\\ e^{i\theta^{(+)}(k)}\end{bmatrix} \right|^{2}\\
&\times&\!\!\!\!|\alpha-\beta-\sqrt{2}\omega\alpha\tilde{f}(e^{i\theta^{(+)}(k)})|^{2}.
\end{eqnarray*}
\noindent
In a similar way, we get the second term of Eq. (\ref{french}) by\\

$\left|Res\left(\dfrac{e^{-ik}}{1-e^{-ik}\tilde{\lambda}^{(-)}(z)}\begin{bmatrix}z\\ \tilde{f}(z)\tilde{\lambda}^{(-)}(z)\end{bmatrix}\left[\dfrac{e^{2\pi i \phi}}{\sqrt{2}},\dfrac{e^{2\pi i \phi}}{\sqrt{2}}\right]\tilde{\Xi}_{0}(z):z=e^{i\theta^{(-)}(k)}\right)\right|^{2}$
\begin{eqnarray*}&=&\!\!\!\!\dfrac{1}{2}\left|Res\left(\dfrac{1}{1-e^{-ik}\tilde{\lambda}^{(-)}(z)}:z=e^{i\theta^{(-)}(k)}\right)\right|^{2}\dfrac{1}{|\tilde{\Lambda}_{0}(e^{i\theta^{(-)}(k)})|^{2}}\left|\begin{bmatrix}e^{i\theta^{(-)}(k)}\\ \tilde{f}(e^{i\theta^{(+)}(k)}) \tilde{\lambda}^{(-)}(e^{i\theta^{(-)}(k)})\end{bmatrix}
\right|^{2}\\&\times&\!\!\!\!|\alpha+\beta-\sqrt{2}\omega\beta\tilde{f}(e^{i\theta^{(-)}(k)})|^{2}.\end{eqnarray*}
\noindent
Consequently, we obtain
\begin{eqnarray}
\|Res(\hat{\tilde{\Xi}}(k:z)\!\!\!\!\!&:&\!\!\!\!\!z=e^{i\theta^{(\pm)}(k)})\|^{2}
=\dfrac{1}{2}\left|Res\left(\dfrac{1}{1-e^{ik}\tilde{\lambda}^{(+)}(z)}:z=e^{i\theta^{(+)}(k)}\right)\right|^{2}\dfrac{1}{|\tilde{\Lambda}_{0}(e^{i\theta^{(+)}(k)})|^{2}}\nonumber\\
&\times&\left|\begin{bmatrix}\tilde{f}(e^{i\theta^{(+)}(k)})\tilde{\lambda}^{(+)}(e^{i\theta^{(+)}(k)})\\ e^{i\theta^{(+)}(k)}\end{bmatrix} \right|^{2}|\alpha-\beta-\sqrt{2}\omega\alpha\tilde{f}(e^{i\theta^{(+)}(k)})|^{2}\nonumber\\
&+&\dfrac{1}{2}\left|Res\left(\dfrac{1}{1-e^{-ik}\tilde{\lambda}^{(-)}(z)}:z=e^{i\theta^{(-)}(k)}\right)\right|^{2}\dfrac{1}{|\tilde{\Lambda}_{0}(e^{i\theta^{(-)}(k)})|^{2}}\nonumber\\
&\times&\left|\begin{bmatrix}e^{i\theta^{(-)}(k)}\\ \tilde{f}(e^{i\theta^{(-)}(k)}) \tilde{\lambda}^{(-)}(e^{i\theta^{(-)}(k)})\end{bmatrix}
\right|^{2}|\alpha+\beta-\sqrt{2}\omega\beta\tilde{f}(e^{i\theta^{(-)}(k)})|^{2}
.\label{kekka}
\end{eqnarray}
\\
Hereafter, we will rewrite the items below in terms of $x_{+}$ or $x_{-}$, and then substitute the items in Eq. \eqref{kekka}.

\begin{enumerate}
\item $\left|Res\left(\dfrac{1}{1-e^{ik}\tilde{\lambda}^{(+)}(z)}: z=e^{i\theta^{(+)}(k)}\right)\right|^{2}$ and $\left|Res\left(\dfrac{1}{1-e^{-ik}\tilde{\lambda}^{(-)}(z)}: z=e^{i\theta^{(-)}(k)}\right)\right|^{2}$,
\item $\dfrac{1}{|\tilde{\Lambda}_{0}(e^{i\theta^{(\pm)}(k)})|^{2}}$,
\item $\dfrac{1}{2|\alpha-\beta-\sqrt{2}\omega\alpha\tilde{f}(e^{i\theta^{(+)}(k)})|^{2}}$ and $\dfrac{1}{2|\alpha+\beta-\sqrt{2}\omega\beta\tilde{f}(e^{i\theta^{(-)}(k)})|^{2}}$,
\item $\left\|\begin{bmatrix}\tilde{\lambda}^{(+)}(e^{i\theta^{(+)}(k)})\tilde{f}(e^{i\theta^{(+)}(k)})\\ e^{i\theta^{(+)}(k)}\end{bmatrix}\right\|^{2}$ and $\left\|\begin{bmatrix}e^{i\theta^{(-)}(k)}\\ \tilde{\lambda}^{(-)}(e^{i\theta^{(-)}(k)})\tilde{f}(e^{i\theta^{(-)}(k)})\end{bmatrix}\right\|^{2}$.\vspace{5mm}
\end{enumerate}
\par\noindent
$1.$ Calculation of $\left|Res\left(\dfrac{1}{1-e^{ik}\tilde{\lambda}^{(+)}(z)}: z=e^{i\theta^{(+)}(k)}\right)\right|^{2}$ and $\left|Res\left(\dfrac{1}{1-e^{-ik}\tilde{\lambda}^{(-)}(z)}: z=e^{i\theta^{(-)}(k)}\right)\right|^{2}$.\\
\noindent
Put $g^{(\pm)}(z)=1-e^{\pm ik}\tilde{\lambda}^{(\pm)}(z)$. Expanding $g^{(\pm)}(z)$ around $z=e^{i\theta^{(\pm)}(k)}$, we get
\[Res\left(\dfrac{1}{1-e^{\pm ik}\tilde{\lambda}^{(\pm)}(z)}:z=e^{i\theta^{(\pm)}(k)}\right)=\left.\dfrac{1}{\dfrac{d g^{(\pm)}(z)}{d z}}\right|_{z=e^{i\theta^{(\pm)}(k)}}.\]
Eq. \eqref{nippon} gives
\[\left.\dfrac{d g^{(\pm)}(z)}{d z}\right|_{z=e^{i\theta^{(\pm)}(k)}}=-\dfrac{\operatorname{sgn}(\cos k)}{\sqrt{1-x_{\pm}^{2}}}e^{-i(\theta^{(\pm)}(k)\mp k)}\left\{1-i\operatorname{sgn}(\cos k\sin k)\dfrac{\sqrt{1-2x_{\pm}^{2}}}{x_{\pm}}\right\},\]
which suggests
\begin{eqnarray}
\left\{
\begin{array}{l}
\left|Res\left(\dfrac{1}{1-e^{ik}\tilde{\lambda}^{(+)}(z)} : z=e^{i\theta^{(+)}(k)}\right)\right|^{2}=x_{+}^{2},\\
\\
\left|Res\left(\dfrac{1}{1-e^{-ik}\tilde{\lambda}^{(-)}(z)}: z=e^{i\theta^{(-)}(k)}\right)\right|^{2}=x_{-}^{2}.
\end{array}
\right.
\end{eqnarray} 
\\ \noindent
$2.$ Calculation of $\dfrac{1}{|\tilde{\Lambda}_{0}(e^{i\theta^{(\pm)}(k)})|^{2}}$.\\
\noindent
From Lemma \ref{prf2.2}, we have 
\begin{align}
&|\tilde{\Lambda}_{0}(e^{i\theta})|^{2}\nonumber\\
&=1+2|\tilde{f}(e^{i\theta})|^{2}+|\{\tilde{f}(e^{i\theta})\}^{2}|-2\sqrt{2}\Re{(e^{2\pi i \phi}\tilde{f}(e^{i\theta}))}+2\Re{(e^{4\pi i \phi}\{\tilde{f}(e^{i\theta})\}^{2})}-2\sqrt{2}\Re{(e^{2\pi i \phi} \overline{\tilde{f}(e^{i\theta})}\tilde{f}(e^{i\theta})^{2})},\nonumber\\
\label{tokyo}
\end{align}
for any real number $\theta\in\mathbb{R}$.
Therefore, substituting the singular points into Eq. \eqref{tokyo}, we obtain
\begin{eqnarray}
\left\{
\begin{array}{l}
\left|\dfrac{1}{\tilde{\Lambda}_{0}(e^{i\theta^{(+)}(k)})}\right|^{2}\\=\dfrac{(1+x_{+})^{2}}{4\{1-\cos(2\pi \phi)+ x_{+}^{2}\cos(4\pi \phi)/2-\operatorname{sgn}(\sin k\cos k)\sin(2\pi \phi)\sqrt{1-2x_{+}^{2}}\sin(2\pi \phi)(1-\cos(2\pi \phi))\}},\\
\\
\left|\dfrac{1}{\tilde{\Lambda}_{0}(e^{i\theta^{(-)}(k)})}\right|^{2}\\=\dfrac{(1-x_{-})^{2}}{4\{1-\cos(2\pi \phi)+ x_{-}^{2}\cos(4\pi \phi)/2+\operatorname{sgn}(\sin k\cos k)\sin(2\pi \phi)\sqrt{1-2x_{-}^{2}}\sin(2\pi \phi)(1-\cos(2\pi \phi))\}}.
\end{array}
\right.
\end{eqnarray} 
\\
\noindent
$3.$ Calculation of $|\alpha-\beta-\sqrt{2}\omega\alpha\tilde{f}(e^{i\theta^{(+)}(k)})|^{2}/2$ and $|\alpha+\beta-\sqrt{2}\omega\beta\tilde{f}(e^{i\theta^{(-)}(k)})|^{2}/2$.\\
\noindent
Put the initial coin state $\Psi_{0}(0)={}^T\![\alpha,\beta]$, where $\alpha=ae^{i\phi_{1}},\;\beta=be^{i\phi_{2}}$ with $a,b\geq0$ and $a^{2}+b^{2}=1$. 
Noting 
\begin{eqnarray*}\dfrac{1}{2}|\alpha\!\!\!&-&\!\!\!\beta-\sqrt{2}\omega\alpha\tilde{f}(e^{i\theta^{(+)}(k)})|^{2}\nonumber\\
\!\!\!&=&\!\!\!\dfrac{1}{2}\{1+2|\alpha|^{2}|\tilde{f}(e^{i\theta^{(+)}(k)})|^{2}-2\Re{(\alpha\overline{\beta})}-2\sqrt{2}|\alpha|^{2}\Re(e^{2\pi i\phi}\tilde{f}(e^{i\theta^{(+)}(k)}))+2\sqrt{2}\Re(\alpha\overline{\beta}e^{2\pi i \phi}\tilde{f}(e^{i\theta^{(+)}(k)}))\},\end{eqnarray*}
and
\begin{eqnarray*}\dfrac{1}{2}|\alpha\!\!\!&+&\!\!\!\beta-\sqrt{2}\omega\beta\tilde{f}(e^{i\theta^{(-)}(k)})|^{2}\nonumber\\
\!\!\!&=&\!\!\!\dfrac{1}{2}\{1+2|\beta|^{2}|\tilde{f}(e^{i\theta^{(-)}(k)})|^{2}+2\Re{(\beta\overline{\alpha})}-2\sqrt{2}|\beta|^{2}\Re(e^{2\pi i\phi}\tilde{f}(e^{i\theta^{(-)}(k)}))-2\sqrt{2}\Re(\overline{\alpha}\beta e^{2\pi i \phi}\tilde{f}(e^{i\theta^{(-)}(k)}))\},\end{eqnarray*}
we obtain
\begin{eqnarray}
\left\{
\begin{array}{l}
\dfrac{1}{2}|\alpha-\beta-\sqrt{2}\omega\alpha\tilde{f}(e^{i\theta^{(+)}(k)})|^{2}\\
=\dfrac{1}{2}+a^{2}\dfrac{1-x_{+}}{1+x_{+}}-ab\cos\tilde{\phi}_{12}-\dfrac{a^{2}}{1+x_{+}}\{\cos(2\pi\phi)+\operatorname{sgn}(\sin k\cos k)\sqrt{1-2x_{+}^{2}}\sin(2\pi\phi)\}\\+
\dfrac{ab}{1+x_{+}}\{\cos(\tilde{\phi}_{12}+2\pi\phi)+\operatorname{sgn}(\sin k\cos k)\sqrt{1-2x_{+}^{2}}\sin(\tilde{\phi}_{12}+2\pi\phi)\},\\
\dfrac{1}{2}|\alpha+\beta-\sqrt{2}\omega\beta\tilde{f}(e^{i\theta^{(-)}(k)})|^{2}\\
=\dfrac{1}{2}+b^{2}\dfrac{1+x_{-}}{1-x_{-}}+ab\cos\tilde{\phi}_{12}-\dfrac{b^{2}}{1-x_{-}}\{\cos(2\pi\phi)-\operatorname{sgn}(\sin k\cos k)\sqrt{1-2x_{-}^{2}}\sin(2\pi\phi)\}\\-
\dfrac{ab}{1-x_{-}}\{\cos(\tilde{\phi}_{21}+2\pi\phi)-\operatorname{sgn}(\sin k\cos k)\sqrt{1-2x_{+}^{2}}\sin(\tilde{\phi}_{21}+2\pi\phi)\}.
\end{array}
\right.
\end{eqnarray} 
\\
\noindent
$4.$ Calculation of $\left\|\begin{bmatrix}\tilde{\lambda}^{(+)}(e^{i\theta^{(+)}(k)})\tilde{f}(e^{i\theta^{(+)}(k)})\\ e^{i\theta^{(+)}(k)}\end{bmatrix}\right\|^{2}$ and $\left\|\begin{bmatrix}e^{i\theta^{(-)}(k)}\\ \tilde{\lambda}^{(-)}(e^{i\theta^{(-)}(k)})\tilde{f}(e^{i\theta^{(-)}(k)})\end{bmatrix}\right\|^{2}$.\\
\noindent
By definition, we have
\begin{eqnarray}
\left\{
\begin{array}{ll}
\left\|\begin{bmatrix}\tilde{\lambda}^{(+)}(e^{i\theta^{(+)}(k)})\tilde{f}(e^{i\theta^{(+)}(k)})\\ e^{i\theta^{(+)}(k)}\end{bmatrix}\right\|^{2}=|\tilde{\lambda}^{(+)}(e^{i\theta^{(+)}(k)})|^{2}|\tilde{f}(e^{i\theta^{(+)}(k)})|^{2}+1=\dfrac{2}{1+x_{+}}&(x_{+}>0),\\
\left\|\begin{bmatrix}e^{i\theta^{(-)}(k)}\\ \tilde{\lambda}^{(-)}(e^{i\theta^{(-)}(k)})\tilde{f}(e^{i\theta^{(-)}(k)})\end{bmatrix}\right\|^{2}=1+|\tilde{\lambda}^{(-)}(e^{i\theta^{(-)}(k)})|^{2}|\tilde{f}(e^{i\theta^{(-)}(k)})|^{2}=\dfrac{2}{1-x_{-}}&(x_{-}<0).
\end{array}
\right.
\end{eqnarray}
\par\noindent
Noting
\begin{align}-\dfrac{d\theta^{(\pm)}(k)}{d k}=x_{\pm},\label{hensu}\end{align}
we see
\begin{align}x_{+}=\dfrac{|\cos k|}{\sqrt{1+\cos^{2}k}},\;\;\;
x_{-}=-\dfrac{|\cos k|}{\sqrt{1+\cos^{2}k}}.\label{x}\end{align}
Thus, we can treat $x_{+}$ and $x_{-}$ as a variable $x$; 
\[
x=\left\{ \begin{array}{ll}
x_{+} & (x>0), \\
x_{-} & (x<0).
\end{array} \right.\] 
\par\noindent
Combining Eqs. (\ref{solutions+}) and (\ref{solutions-}) with Eq. (\ref{x}), and noting Eq. (\ref{hensu}), we have
\[\dfrac{dx}{dk}=\mp\operatorname{sgn}(\sin k\cos k)(1-x^{2})\sqrt{1-2x^{2}}.\]
Hence, we see
\begin{align}
dk=\left\{ \begin{array}{ll}
-\operatorname{sgn}(\sin k\cos k)f_{K}(x;1/\sqrt{2})\pi dx&(x>0), \\
\operatorname{sgn}(\sin k\cos k)f_{K}(x;1/\sqrt{2})\pi dx& (x<0).
\end{array} \right.
\end{align} 
\noindent
Substituting the items given in $1.$ to $4.$ into Eq. \eqref{kekka} and combining with Eq. \eqref{limitdensity}, the proof of Theorem \ref{weaklimit} is completed.
\vspace{5mm}\\
\noindent {\large{\bf Appendix B}}  \\
\noindent
We give the detailed explanation of Eq. \eqref{limitdensity}, which is a key to prove Theorem \ref{weaklimit}. Put
$\omega_{l}(k)=Res(\hat{\tilde{\Psi}}_{t}(k:z): z=e^{i\theta^{(l)}(k)})$ with $\Psi_{t}(x)=\Xi_{t}(x)\varphi_{0}$ and $l=+, -$. Then, we have by definition
\begin{eqnarray}E\left[e^{i\xi \frac{X_{t}}{t}}\right]\!\!&=&\!\!\sum_{j}P(X_{t}=j)e^{i\xi\frac{j}{t}}\nonumber\\
\!\!\!&=&\!\!\!\sum_{j}\|\Xi_{t}(j)\varphi_{0}\|^{2}e^{i\xi\frac{j}{t}}\nonumber\\
\!\!\!&=&\!\!\!\int^{2\pi}_{0}\sum_{x,y}\varphi_{0}^{\ast}\Xi_{t}^{\ast}(y)\Xi_{t}(x)\varphi_{0}e^{i\xi\frac{x}{t}}e^{ik(x-y)}\dfrac{dk}{2\pi} \nonumber\\
\!\!&=&\!\!\int^{2\pi}_{0}\sum_{x,y}\left<\Psi_{t}(y), \Psi_{t}(x)\right>e^{i\xi\frac{x}{t}}e^{ik(x-y)}\dfrac{dk}{2\pi}\nonumber\\
\!\!\!&=&\!\!\!\int^{2\pi}_{0}\left<\hat{\Psi}_{t}(k), \hat{\Psi}_{t}\left(k+\dfrac{\xi}{t}\right)\right>\dfrac{dk}{2\pi}\nonumber\\
\!\!\!&=&\!\!\!\int^{2\pi}_{0}\left<\sum_{l}\omega_{l}(k)e^{-i(t+1)\theta^{(l)}(k)},\sum_{m}\omega_{m}\left(k+\dfrac{\xi}{t}\right)e^{-i(t+1)\theta^{(m)}(k+\frac{\xi}{t})}\right>\dfrac{dk}{2\pi}\label{ringo}\\
\!\!\!&=&\!\!\!\int^{2\pi}_{0}\left\{\sum_{l}|\omega_{l}(k)|^{2}e^{-i\xi\frac{t+1}{t}\frac{d\theta^{(l)}(k)}{d k}} e^{-i(t+1)O\left(\frac{1}
{t^{2}}\right)}+O\left(\frac{1}{t}\right)\right\}\dfrac{dk}{2\pi}\nonumber\\
\!\!\!&+&\!\!\!\int^{2\pi}_{0} \left\{\sum_{l}\sum_{m}\!\!\overline{\omega_{l}(k)}e^{i(t+1)\theta^{(l)}(k)}\omega_{m}(k)e^{-i(t+1)\theta^{(m)}(k)}e^{-i\xi\frac{t+1}{t}\frac{d \theta^{(m)}(k)}{d k}} e^{-i(t+1) O\left(\frac{1}{t^{2}}\right)}+O\left(\frac{1}{t}\right)\right\}\dfrac{dk}{2\pi}.\nonumber\\
\label{syomei}
\end{eqnarray}
We took advantage of the residue theorem \[\int^{2\pi}_{0}\hat{\tilde{\Psi}}_{t}(k;z)dz=2\pi i\sum_{i}Res(\hat{\tilde{\Psi}}_{t}(k;z),\;z=\theta^{(i)}(k))\] and the inverse Fourier transform \[\hat{\Psi}_{t}(k)=\frac{1}{2\pi i}\int^{2\pi}_{0}\hat{\tilde{\Psi}}_{t}(k;z)\frac{dz}{z^{t+1}}\] to obtain Eq. \eqref{ringo}. Using Maclaurin's expansion for $w_{m}(k+\xi/t)e^{-i(t+1)\theta^{(m)}(k+\xi/t)}$, that is,
\[w_{m}\left(k+\xi/t\right)e^{-i(t+1)\theta^{(m)}(k+\xi/t)}=\left(w(k)+\dfrac{\xi}{t}\dfrac{d w(k)}{d k}+\dfrac{\xi^{2}}{2t^{2}}\dfrac{d^{2} w(k)}{d^{2} k}+\cdots\right)e^{-i(t+1)\left\{\theta^{(m)}(k)+\frac{\xi}{t}\frac{d \theta^{(m)}(k)}{d k}+\frac{\xi^{2}}{2t^{2}}\frac{d^{2}\theta^{(m)}(k)}{d^{2} k}+\cdots\right\}},\]
we got Eq. \eqref{syomei}.
According to the Riemann-Lebesgue Theorem, the second term of Eq. (\ref{syomei}) vanishes when $t\to\infty$, and we obtain the desired relation.
\vspace{5mm}\\
\noindent {\large{\bf Appendix C}}  \\
\noindent
Hereafter, we explain in detail, how $\tilde{f}(z)$ and $\tilde{\lambda}^{(\pm)}(z)$ are determined when we focus on the ballistic behavior of the Wojcik model. 
Owing to \cite{segawa}, we see
\[
\left\{
\begin{array}{l}
\tilde{\lambda}^{(\pm)}(\omega)=\pm\dfrac{i}{\sqrt{2}}\{(\omega+\omega^{-1})-\sqrt{(\omega+\omega^{-1})^{2}-2}\},\\
\tilde{f}(\omega)=-\dfrac{\omega }{\sqrt{2}}\{(\omega-\omega^{-1})+\sqrt{(\omega+\omega^{-1})^{2}-2}\}.
\end{array}
\right.\]
Putting $\omega=i(1-\epsilon)e^{i\theta}\;(\epsilon\in\mathbb{R},\;|\epsilon|\ll1)$, we consider how $\lim_{\epsilon\to0}\sqrt{(\omega+\omega^{-1})^{2}-2}$ can be written in terms of $\theta$ with the range of $\cos\theta$ or $\sin\theta$.
Note $|\epsilon|\ll1$, and we can approximate $\tilde{\lambda}^{(\pm)}(\omega)$ as \cite{endosan}
\begin{eqnarray}
\tilde{\lambda}^{(\pm)}(\omega)\!\!\!&=&\!\!\!\pm\dfrac{i}{\sqrt{2}}\left\{(1-\epsilon)ie^{i\theta}-(1-\epsilon)^{-1}ie^{-i\theta}-\sqrt{\{(1-\epsilon)ie^{i\theta}-(1-\epsilon)^{-1}ie^{-i\theta}\}^{2}-2}\right\}\nonumber\\
\!\!\!&\sim&\!\!\!\mp\dfrac{i}{\sqrt{2}}\left\{2\sin\theta+2i\epsilon\cos\theta+\delta\sqrt{4\sin^{2}\theta-2}\right\},\label{gyoten}\end{eqnarray}
where we put $\delta\in\mathbb{R}$ with $\delta^{2}=1$. 
Noting $|\tilde{\lambda}^{(\pm)}(\omega)|<1$,
Eq. \eqref{gyoten} suggests that we need to take into account the next two cases.
\begin{enumerate}
\item Case of $|\sin\theta|\geq 1/\sqrt{2}$:\\
Eq. \eqref{gyoten} gives  
\[\dfrac{1}{2}\left\{2\sin\theta+2\delta\sqrt{\sin^{2}\theta-\dfrac{1}{2}}\right\}^{2}<1.\]
Hence, we have
\[2\sin^{2}\theta+2\sin\theta\delta\sqrt{\sin^{2}\theta-\dfrac{1}{2}}<1.\]
Consequently, we get $\delta=-\operatorname{sgn}(\sin\theta)$.
\item Case of $|\sin\theta|<1/\sqrt{2}$:\\
Eq. \eqref{gyoten} also gives 
\[\dfrac{1}{2}\left[4\sin^{2}\theta+\left\{2\epsilon\cos\theta+2\delta\sqrt{\dfrac{1}{2}-\sin^{2}\theta}\right\}^{2}\right]<1.\]
Therefore, we see
\[4\epsilon^{2}\cos^{2}\theta+8\epsilon\cos\theta\delta\sqrt{\dfrac{1}{2}-\sin^{2}\theta}<0.\]
Consequently, we obtain $\delta=-\operatorname{sgn}(\cos\theta)$.
\end{enumerate}
Accordingly, the square root is expressed as
\begin{align}\lim_{\epsilon\to0}\sqrt{(\omega+\omega^{-1})^{2}-2}=\left\{
\begin{array}{ll}
-2\operatorname{sgn}(\sin\theta)\sqrt{\sin^{2}\theta-\dfrac{1}{2}}&(\;|\sin\theta|\ge1/\sqrt{2}\;),\\
-2i\operatorname{sgn}(\cos\theta)\sqrt{\dfrac{1}{2}-\sin^{2}\theta}&(\;|\sin\theta|<1/\sqrt{2}\;).
\end{array}
\right.\label{news}\end{align}

Next, we determine in detail $\tilde{\lambda}^{(\pm)}(z)$ and $\tilde{f}(z)$.
When we consider the weak convergence theorem for our Wojcik model, we choose the square root so that $1/(1-e^{ik}\tilde{\lambda}^{(+)}(z))$ and $1/(1-e^{-ik}\tilde{\lambda}^{(-)}(z))$ have the singular points, that is, $|\tilde{f}(z)|\neq1$.
Therefore, we see from Eqs. \eqref{gyoten} and \eqref{news}, 
\[\left\{
\begin{array}{l}
\tilde{\lambda}^{(\pm)}(z)=\mp\{\operatorname{sgn}(\cos\theta)\sqrt{2\cos^{2}\theta-1}+i\sqrt{2}\sin\theta\},\\
\\
\tilde{f}(z)=\operatorname{sgn}(\cos\theta)e^{i\theta}\{\sqrt{2}|\cos\theta|-\sqrt{2\cos^{2}\theta-1}\},
\end{array}
\right.\;(|\sin\theta|<1/\sqrt{2})
\]
with $z=e^{i\theta}$.

\end{document}